\documentclass[a4paper,fleqn]{cas-sc}

\usepackage[numbers]{natbib}
\usepackage{enumitem}
\setlist{nosep}
\usepackage{tikz}
\usepackage{placeins}
\usetikzlibrary{arrows.meta,positioning}

\newcommand{\chfour}{CH\textsubscript{4}}
\newcommand{\cotwo}{CO\textsubscript{2}}
\newcommand{\pp}{\,p.p.}

\ExplSyntaxOn
\cs_set:Npn \__first_footerline:
{
  \group_begin:
  \small
  \sffamily
  \ifnum\theblind>0\relax
  \else
    \__short_authors:
  \fi
  \group_end:
}
\ExplSyntaxOff

\makeatletter
\AtEndDocument{\clearpage\immediate\write\@auxout{%
  \string\csxdef{lastpage}{\the\numexpr\value{page}-1\relax}}}
\makeatother

\begin{document}
\let\WriteBookmarks\relax
\def\floatpagepagefraction{1}
\def\textpagefraction{.001}

% =============================================================================
% FRONT MATTER
% =============================================================================
\shorttitle{RTO with online learning for PSA}
\shortauthors{C. de Menezes Rebello et al.}

\title[mode=title]{Real-time optimization with online learning for pressure
swing adsorption}

\author[1]{Carine de Menezes Rebello}[orcid=0000-0002-0796-8116]
\cormark[1]
\ead{carine.m.rebello@ntnu.no}

\author[2]{Marcio Fontana}[orcid=0000-0001-9917-5887]

\author[3]{Leizer Schnitman}[orcid=0000-0002-0399-6689]

\author[1,4]{Idelfonso~B.~R. Nogueira}[orcid=0000-0002-0963-6449]

\affiliation[1]{organization={Puffin HUB},
    city={Trondheim},
    country={Norway}}

\affiliation[2]{organization={School of Electrical and Computer Engineering,
    University of Campinas (UNICAMP)},
    addressline={Avenida Albert Einstein 400},
    postcode={13083-852},
    city={Campinas},
    state={SP},
    country={Brazil}}

\affiliation[3]{organization={Graduate Program in Mechatronics, Federal
    University of Bahia (UFBA)},
    addressline={Rua Prof. Aristides Novis 2},
    postcode={40210-630},
    city={Salvador},
    state={BA},
    country={Brazil}}

\affiliation[4]{organization={Department of Chemical Engineering, Norwegian
    University of Science and Technology (NTNU)},
    city={Trondheim},
    country={Norway}}

\cortext[cor1]{Corresponding author}

\begin{abstract}
Real-time optimization (RTO) of pressure swing adsorption (PSA) units requires
a process model that can be evaluated many times at each decision, which
favours data-driven surrogates of the cyclic steady state. A surrogate,
however, represents the plant only under the conditions covered by its
training data, and disturbances that are not among its inputs remain outside
its description. This work presents an RTO framework for PSA in which a
neural network surrogate, optimized by particle swarm, is updated online with
data generated at the plant when plant and model diverge. A rigorous dynamic
model of a \chfour/\cotwo{} separation, integrated to cyclic steady state,
served as virtual plant in three scenarios. In the first, five consecutive RTO events
with different objectives ran without disturbance; all constraints were
satisfied and the largest deviation between surrogate and plant was 1.7
percentage points. In the second, a step in the \cotwo{} feed fraction from
0.40 to 0.50 reduced the \chfour{} purity at the operating point in force from 92.3 to
76.9\,\%; after the network was fine-tuned on 43 local samples, the new RTO event
returned a feasible optimum with a purity of 90.3\,\%. In the third, a progressive loss of sorbent
capacity was detected by the accumulated relative deviation of the \chfour{}
purity from a reference, triggering retraining before any constraint
violation. Samples generated across the remaining activity range led to
infeasible operation, whereas samples generated at the final activity kept the
purity above 92.4\,\% after the move transient.
\end{abstract}

\begin{highlights}
\item Real-time optimization of PSA with a neural surrogate retrained online.
\item Five sequential objectives met with surrogate errors of at most 1.7 p.p.
\item Retraining triggered by the accumulated deviation of the plant from a reference.
\item A disturbance absent from the surrogate inputs was corrected with 43 local samples.
\item Under a progressive disturbance, data at its final value kept the optimum feasible.
\end{highlights}


\maketitle

% =============================================================================
\section{Introduction}\label{sec:intro}
% =============================================================================
Industrial plants undergo many changes during operation, of different
origins: the feed varies in flow rate and composition, both as an uncertain
parameter that process design must accommodate \cite{halemane1983,swaney1985}
and with the use of new feedstocks of variable composition \cite{mitsos2018};
energy prices fluctuate with the growing share of renewable electricity
\cite{mitsos2018}; and equipment and materials degrade, as in the deactivation
of catalysts \cite{bartholomew2001}, the fouling of heat transfer surfaces
\cite{mullersteinhagen2011} and the loss of capacity of aged adsorbents
\cite{nascimento2021}. An operating point that was optimal at
commissioning becomes suboptimal, and may become infeasible, as these
conditions drift \cite{skogestad2000,jaschke2017}. The difference between the
performance obtained at the current operating point and the performance
attainable under the current conditions is an economic margin that is lost
unless the operating point is updated \cite{skogestad2000}, and recovering it
is the purpose of the optimization layers of process automation
\cite{naysmith1995,engell2007,tatjewski2008,darby2011,camara2016}.

Real-time optimization (RTO) computes the operating point at regular
intervals by optimizing a
steady-state model of the process, which is updated with plant measurements
before each optimization \cite{darby2011,camara2016}. The result is only as
good as this model. A model-based RTO system reaches the plant optimum only if
the model is adequate, that is, if it can predict an optimum at the same point
as the plant \cite{forbes1994}, and when the structure of the model differs
from that of the plant, updating its parameters does not guarantee this
\cite{chachuat2009}. This motivated modifier adaptation and related schemes,
which correct the optimization problem with plant information
\cite{marchetti2009,marchetti2016}. Steady-state RTO must also wait for the
plant to settle after each move and after each disturbance before the model is
updated, which limits how fast it follows the changes \cite{krishnamoorthy2018}.

The update of the model becomes a different problem when the model is
data-driven. Machine learning has been used in chemical engineering to build
process models, alone \cite{rebello2022} or combined with first-principles
balances in hybrid models \cite{lima2023}, and to design molecules
\cite{queiroz2023omega,queiroz2023foods}. In RTO, such models have been used
either to represent the plant-model mismatch with Gaussian processes
\cite{ferreira2018,delriochanona2021} or as the process model itself
\cite{ehlhardt2023,costa2026}; RTO based on surrogate models was shown to
re-optimize an electric submersible pump after unmeasured disturbances at a
fraction of the time required by the first-principles model \cite{costa2026}.
A data-driven model, however, represents the process only within the
conditions covered by its training data, and its accuracy decreases outside
them, as shown for adsorption processes \cite{rebello2022}. When the plant
changes, such a model is not corrected by adjusting a few physical parameters
but by new data. The adaptation of data-driven models to changes in the process
has been studied for soft sensors \cite{kadlec2011}, for model predictive
control with online learning and transfer learning of neural networks
\cite{hu2023,alhajeri2024}, and in digital twin frameworks with online learning
\cite{rebello2025dt,rebello2025ifac}, one of which follows the behaviour of an
adsorption unit, including adsorbent degradation \cite{costa2024}, as part of the wider integration of
artificial intelligence and cyber-physical systems in the chemical industry
\cite{oliveira2021ai} and of sensor networks and data analytics in utility
systems \cite{santossanchez2020}.

Pressure swing adsorption (PSA) is a demanding application of these ideas,
because of both the variability of its operating conditions and the complexity
of its models. In PSA, adsorbent beds are cycled between a high pressure, at
which the more strongly adsorbed components are retained, and a low pressure,
at which they are released \cite{ruthven1994,sircar2002}. In operation, feed
composition varies, and the response of a PSA unit to feed
disturbances has been studied experimentally \cite{santos2013} and through
the optimization of the unit for variable feed composition and flow rate
\cite{hasan2012,kim2015}. Adsorbent
capacity also changes over time: water reduces \cotwo{} adsorption on
zeolites \cite{brandani2004,li2008}, and impurities such as H\textsubscript{2}S
affect adsorbent performance and regeneration over repeated cycles
\cite{alonsovicario2010}. Both effects change, during the life of the unit,
the conditions under which the product specification must be met.

Identifying the optimal operating point of a PSA unit is not
straightforward. Its operation is defined by a set of coupled variables, such
as step durations, pressures and the fractions of the internal streams recycled
between steps, whose effects interact \cite{ruthven1994,sircar2002}, as also
shown for moving bed separations \cite{nogueira2016,nogueira2017cace}. Its
performance is measured by indicators that conflict with each other: product purity and recovery,
productivity and energy consumption \cite{haghpanah2013iecr,subraveti2019}.
For this reason, PSA design and operation have been treated as optimization
problems for more than three decades
\cite{smith1991,biegler2004,agarwal2010,haghpanah2013iecr}, from the
formulation of the optimal design \cite{smith1991} and of the optimization of
periodic adsorption processes \cite{nilchan1998} to the simultaneous solution of
the cyclic steady state (CSS) and the optimization problem \cite{jiang2003,ko2003,ko2005,biegler2004}
and the multi-objective optimization of the purity-recovery and
energy-productivity trade-offs
\cite{haghpanah2013iecr,haghpanah2013aiche,capra2018}. These studies were
carried out offline, for fixed feed conditions and adsorbent properties, with
rigorous models whose computational cost is the main obstacle to their use
during operation.

A mechanistic PSA model couples mass,
energy and momentum balances with nonlinear adsorption equilibria and kinetics
in a system of partial differential and algebraic equations, whose solution
develops steep concentration and temperature fronts that require fine
discretization and stiff integrators \cite{biegler2004,haghpanah2013iecr}.
Moreover, the performance indicators are defined at CSS, which is reached only
after the integration of the model over tens to hundreds of cycles. Methods to
accelerate the convergence to CSS \cite{smith1992,ding2001} or to determine it
directly within the optimization problem \cite{jiang2003,latifi2008} reduce
this burden, but each evaluation of the model still requires the solution of a
large dynamic problem, and the optimization of PSA units for new adsorbents
and separations \cite{yun2023,santana2024}, including the simultaneous
selection of adsorbents and operating conditions \cite{santana2026}, relies on
many such evaluations. For the optimization of a single cycle configuration
for a single adsorbent, a cost of the order of 1500 core-hours with the
detailed model has been reported \cite{pai2020iecr}. An RTO layer, which must
evaluate the model many times at each decision while the plant is in
operation, cannot afford this cost.

This limitation has led to the use of surrogate models
\cite{bhosekar2018,mcbride2019}, which approximate the map between operating
variables and CSS indicators at a small fraction of the cost of the
mechanistic model \cite{leperi2019,pai2020iecr}. Early approaches used reduced-order models
obtained by proper orthogonal decomposition \cite{agarwal2009} and kriging
\cite{beck2015,alkebsi2021}, but over the last years artificial neural networks
(ANN) have become the most frequent choice in the PSA literature. ANN
surrogates trained on rigorous simulations have been used to optimize PSA
units for methane-nitrogen separation \cite{santanna2017}, hydrogen
purification \cite{ye2019,xiao2020,tong2021,yu2021} and \cotwo{} capture
\cite{leperi2019,subraveti2019,pai2020sep}, and for olefin/paraffin
separation by vacuum PSA \cite{lim2025}, and the reported reductions in
computational effort range from about one order of magnitude in the
multi-objective optimization of a cycle \cite{subraveti2019} to the
replacement of the core-hour budget mentioned above by minutes
\cite{pai2020iecr}. The approach has been extended to networks that include
the residual of the model equations in their training \cite{subraveti2022,wu2025},
to hybrid models that combine adsorption balances with neural networks
\cite{nogueira2022cjce,santana2023,gama2024} and to symbolic regression surrogates
\cite{rebello2026sct},
and Bayesian optimization with Gaussian process surrogates and hybrid
strategies that alternate surrogate and rigorous evaluations have been
proposed to further reduce the number of simulations
\cite{hao2021,ward2022,ma2026,stander2025}. Our group has used deep neural
networks to model PSA dynamics \cite{oliveira2020,rebello2022} and as
surrogates for the optimization of PSA units \cite{rebello2025cjce},
including the mapping of
uncertainty in the optimum \cite{nogueira2020}, the definition of feasible
operating regions and Pareto regions \cite{rebello2021b,rebello2022cherd,rebello2024}
and the simultaneous screening of adsorbents and process optimization
\cite{nogueira2022cherd}, including electric swing adsorption
\cite{regufe2021}, and deep neural network surrogates were
also used for the optimization of a simulated moving bed reactor
\cite{santana2023omega}.

Like the optimization studies that preceded them, these surrogates are trained
once and used offline, so that a change in feed composition or adsorbent
capacity is outside what they represent, which is the limitation of
data-driven models discussed above.

For PSA, the online use of models has been directed mainly to control. Model
predictive controllers have been developed to track product purity and
recovery, with explicit multi-parametric formulations \cite{khajuria2011} also
combined with the optimization of the unit \cite{khajuria2013}, for rapid PSA
oxygen concentrators \cite{urich2018,urich2022}, with hybrid systems theory
\cite{fakhroleslam2019}, with robust formulations
\cite{oliveira2021ifac,martins2023} and in embedded implementations
\cite{mendes2024}; in all of them the set-points are defined by another layer.
An economic nonlinear model predictive controller based on artificial
intelligence models was applied to a PSA unit for syngas purification, without
reported updating of the model \cite{martins2021}. The real-time optimization of PSA units, in contrast, has received little
attention. The only work found that treats it explicitly applies a model-free
extremum seeking scheme, in which the operating point is adjusted with
gradients estimated from plant measurements, without a process model, and its
authors describe the RTO of PSA as an underdeveloped subject
\cite{dasilva2025}.

This body of work reveals a relevant gap: an adaptive RTO for PSA, able to
update its model when the process conditions change. The surrogates used to
optimize PSA units are trained offline, the models used online serve control,
and, to the authors' knowledge, no work has coupled the online updating of a
model to the re-optimization of the operating point of a PSA unit. A surrogate
whose inputs do not include the disturbing variable cannot be corrected by
re-estimating a few parameters; it must be retrained with new data from the
plant. The adaptive RTO of PSA is therefore an open opportunity for research.

Developing it raises two questions that are specific to cyclic adsorption
processes and have no answer in the literature. The first is when the surrogate should be
retrained, given that a slow change in the plant takes many cycles to become
apparent at CSS and that a criterion based on constraint violation acts only
after the specification has been lost. The second is at which plant condition
the new data should be generated, since under a progressive disturbance the
condition at which the data are collected is not the condition under which the
new operating point will be applied.

The objective of this work is to develop an adaptive RTO framework for PSA
units, in which a neural network surrogate of the CSS indicators is optimized
by particle swarm and is retrained online when the plant departs from it, and
to answer the two questions above within this framework. Retraining is
triggered by the accumulated deviation of a plant indicator from the value
reached after the last RTO event, which detects a slow drift before a
constraint is violated, and the new data are generated by local samples at the
plant, with two choices of the plant condition compared under a progressive
disturbance.

The framework is applied to the separation of \chfour{} and \cotwo{} by PSA,
with a rigorous dynamic model integrated to CSS as virtual plant, in three
scenarios. The first applies a sequence of five RTO events with different
objectives and constraint sets to the undisturbed plant and compares the
surrogate predictions with the plant. The second applies a step in feed
composition, a disturbance that is not an input of the surrogate. The third
applies a progressive loss of sorbent capacity and compares the two choices of
the condition at which the retraining data are generated.

The paper is organized as follows. Section~\ref{sec:met} presents the
methodology: the acquisition and processing of data, the identification of the
neural network, its retraining and the optimization. Section~\ref{sec:case}
describes the separation used as case study, with the rigorous model given in
Appendix~\ref{sec:model}. Section~\ref{sec:res} reports the identified
network and the three scenarios, and Section~\ref{sec:conc} summarizes the
conclusions.

% =============================================================================
\section{Methodology}\label{sec:met}
\begin{figure}[!htbp]
\centering
\includegraphics[width=\linewidth]{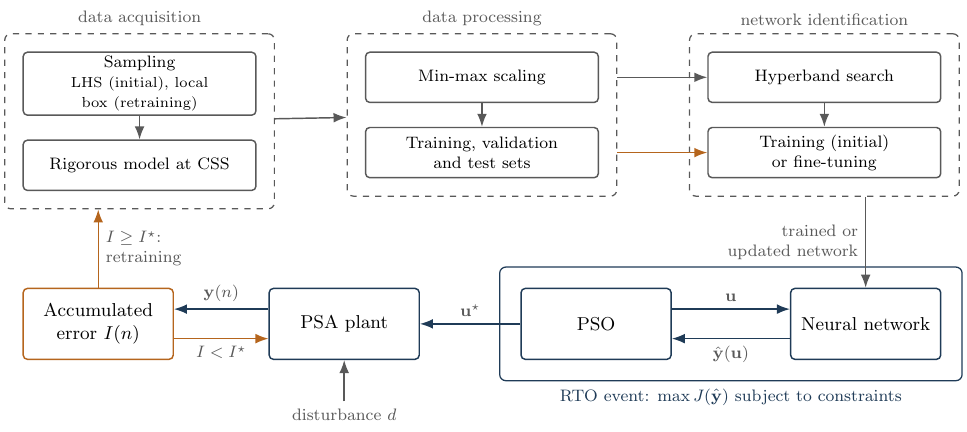}
\caption{Proposed methodology: construction of the neural network surrogate and adaptive RTO.}
\label{fig:framework}
\end{figure}

% =============================================================================
The proposed methodology is summarized in Fig.~\ref{fig:framework}. It
combines a neural network surrogate of the plant indicators at CSS, built
offline from process data, with an RTO layer that optimizes
the surrogate and a retraining procedure that updates the network online when
the plant departs from it. Section~\ref{sec:aq} describes the acquisition and processing of
data: the data sources, the sampling of the inputs of a rigorous model by an
optimized Latin hypercube design, and the treatment, scaling and division of
the data into training, validation and test sets. Section~\ref{sec:surr}
presents the identification of the neural network, whose architecture is
selected by Hyperband search before its weights are trained. Section~\ref{sec:ol} describes
the retraining of the network: the trigger based on the accumulated deviation
of the plant from a reference, the generation of local samples around the
operating point in force and the fine-tuning of the network from its current
weights, without a new architecture search. Section~\ref{sec:rto} presents the
optimization problem solved at each RTO event, the particle swarm optimizer and
the implementation of the optimum at the plant.

% -----------------------------------------------------------------------------
\subsection{Data acquisition and processing}\label{sec:aq}
% -----------------------------------------------------------------------------
The data used to build the neural network must cover the operating limits of
the process. Several data sources can be used, alone or
combined, among them historical process data and synthetic data obtained from
simulations of a rigorous model of the process. Synthetic data can also be
generated in specific scenarios, for instance at operating conditions or
process conditions that are rare or absent in the historical data. When a
rigorous model is used, its inputs $\mathbf{u}$ are varied over their
admissible ranges in order to perturb the mechanistic model and record its
response across the operating domain. Each set of input values is applied to
the model as a step change of the operating point, starting from the state
left by the previous one, and the model is integrated until CSS is reached;
each simulation that reaches CSS gives one data point, formed by the model
inputs and the indicators at CSS.

Because each data point requires one such simulation, the number of points
that can be generated is limited, and the values given to the inputs determine
how well this limited set of samples represents the operating domain. Choosing these values is a design of experiments problem, and the
available designs differ in how they distribute the samples over the domain
\cite{garud2017}. Full factorial grids \cite{montgomery2017} cover the domain
regularly, but the number of points grows exponentially with the number of
variables, so that ten levels of five variables already require $10^5$
simulations. Random (Monte Carlo) sampling removes this dependence on the
number of variables but, with a small number of samples, leaves regions of the
domain without samples and groups samples in others
\cite{niederreiter1992,helton2003}. Quasi-random sequences,
such as the Sobol sequence \cite{sobol1967}, and Latin hypercube designs
\cite{mckay1979} address both limitations, covering the domain more uniformly
than random sampling with a number of samples that is chosen independently of
the number of variables. Among these designs, the methodology adopts an
optimized Latin hypercube design. Latin hypercube sampling divides the range of
each variable into intervals of equal probability and takes one sample in each,
and its optimized forms spread the samples over the domain and keep the
correlation between the inputs low \cite{iman1982,morris1995}.

\begin{figure}[!htbp]
\centering
\includegraphics[width=0.9\linewidth]{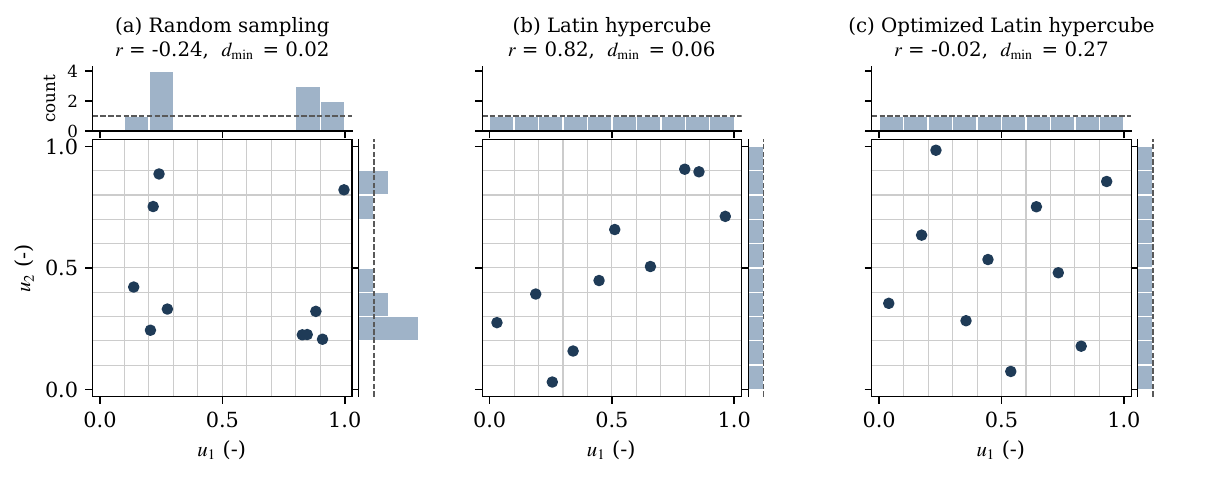}
\caption{Sampling of two normalized variables with ten samples: (a) random,
(b) Latin hypercube and (c) optimized Latin hypercube. The histograms give the
number of samples in each interval of each variable, with the dashed line at
one sample per interval; $r$ is the correlation coefficient between the two
variables and $d_{\min}$ the smallest distance between two samples.}
\label{fig:lhs}
\end{figure}

Fig.~\ref{fig:lhs} compares three ways of drawing ten samples of two
normalized variables. Random sampling (Fig.~\ref{fig:lhs}a) groups the samples,
and six of the ten intervals of $u_1$ and five of $u_2$ receive no sample.
Latin hypercube sampling (Fig.~\ref{fig:lhs}b and c) places one sample in each
interval of each variable, as the flat histograms show, but this does not
prevent the samples from aligning: in Fig.~\ref{fig:lhs}b the correlation
between $u_1$ and $u_2$ is 0.82. The optimized design (Fig.~\ref{fig:lhs}c)
spreads the samples over the domain, with a correlation of $-0.02$ and a
smallest distance between samples of 0.27, against 0.02 and 0.06 in
Fig.~\ref{fig:lhs}a and b. Inputs that vary independently allow the network to
distinguish the effect of each variable on the indicators, and a uniform
coverage keeps its accuracy similar over the region explored by the optimizer.
Historical process data, in contrast, often have strongly correlated inputs,
since the operating variables are moved together by the control system and by
operating practice \cite{kadlec2009}. For this reason, combining different
data sources, with synthetic data generated by a sampling design complementing
the historical data, is a useful strategy to cover the operating domain.

The data are then processed before training. Historical data require a
treatment of missing values, noise and outliers, and of variables recorded
with different sampling times, which must be brought to a common time base;
since the surrogate represents the indicators at CSS, only the periods in
which the plant is at CSS are retained \cite{kadlec2009,cao1995}.
Inputs and outputs are scaled to $[0,1]$ by min-max scaling. The variables of
a process have different units and orders of magnitude, and without scaling
those with larger values dominate the loss function and the updates of the
weights, while large inputs drive the activation functions to saturation and
slow down training; scaling gives all inputs a comparable influence at the
start of training and all outputs a comparable weight in the loss
\cite{lecun1998}. The data points are then divided at random into training, validation and test subsets in
proportions of 70, 15 and 15\,\%.

% -----------------------------------------------------------------------------
\subsection{Network identification}\label{sec:surr}
% -----------------------------------------------------------------------------
The identification of a neural network comprises two stages. The first
selects the hyperparameters, which define the structure of the network and the
conditions of its training and must be fixed before training. The second
estimates the parameters, the weights and biases, by training the network with
the selected hyperparameters. In the methodology, the identified network is a
feed-forward network with multiple inputs and multiple outputs, which maps the
inputs $\mathbf{u}$, together with any fixed input of the process, to all the
indicators at CSS. Such networks can approximate continuous maps on a bounded
domain to any accuracy \cite{hornik1989}, and a single network for all the
indicators requires one evaluation for each candidate point examined by the
optimizer.

The hyperparameters, such as the number of hidden layers and of neurons in
each layer, the activation function, the regularization and the learning rate,
determine the capacity of the network and the ease of its training. A network
with too few layers or neurons cannot represent the map between inputs and
indicators, while an excessively large one requires more data and tends to fit
features of the training points that do not hold elsewhere in the domain. The
suitable combination depends on the process and on the data and cannot be
fixed in advance, and choosing it by trial and error makes the result depend
on the experience of the user, so a systematic search over a defined space is
used. Since each configuration examined requires the training of a network,
the methods differ mainly in how they spend this effort. Grid search becomes
impractical as the number of hyperparameters grows. Random search examines
more values of each hyperparameter for the same number of trainings
\cite{bergstra2012}, but trains every configuration to the end.

\begin{figure}[!htb]
\centering
\includegraphics[width=\linewidth]{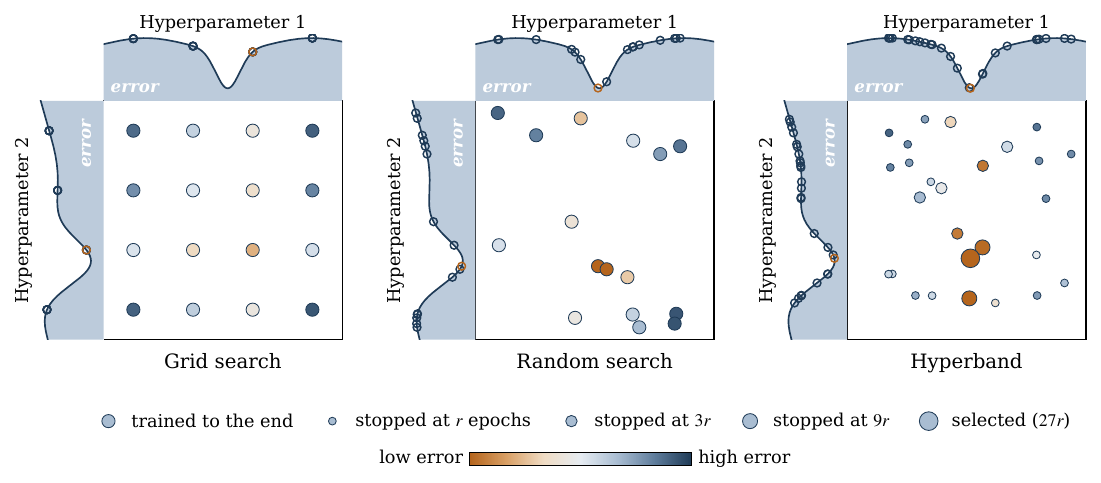}
\caption{Grid search, random search and Hyperband in a space of two
hyperparameters, with the validation error along each hyperparameter shown on
the edges (illustrative values). In Hyperband, the marker size indicates the
training epochs received by each configuration.}
\label{fig:busca}
\end{figure}

Hyperband \cite{li2018hyperband}, adopted here, allocates the training effort
adaptively. Candidate configurations are drawn at random from the search space
and trained for a few epochs, and in successive rounds only the best $1/\eta$
of them, according to the validation error, continue with $\eta$ times more
epochs, where $\eta$ is the reduction factor, up to a maximum number of epochs.
The ranking may change as training proceeds, and a
configuration that learns slowly may be discarded early although it would
perform well with longer training. For this reason, several such sequences are
run, starting with fewer candidates and more epochs, which balances the number
of configurations examined against the effort spent on each. Poor configurations
are thus
discarded after a few epochs, and more configurations are examined for the
same computational effort than when each is trained to the end
\cite{li2018hyperband}. Fig.~\ref{fig:busca} compares the three strategies in
a space of two hyperparameters: the grid fixes the values in advance and can
miss a narrow region of low error, random search covers more values of each
hyperparameter, and Hyperband samples as random search but extends the training
only of the configurations with lower error. The search space is given in Table~\ref{tab:hpspace}:
the candidates differ in the number of hidden layers, the number of neurons in
each layer, the activation function, the use of batch normalization of the
layer outputs \cite{ioffe2015} and of dropout \cite{srivastava2014}, and the
learning rate. The objective of the search is the mean absolute error on the
validation subset.

\begin{table}[!htb]
\centering
\caption{Search space of the Hyperband search.}
\label{tab:hpspace}
\small
\begin{tabular}{ll}
\toprule
Hyperparameter & Values \\
\midrule
Hidden layers & 1 to 5 \\
Neurons per hidden layer & 64 to 512, step 64 \\
Activation function & ReLU, hyperbolic tangent, ELU \\
Batch normalization & yes, no \\
Dropout & yes, no \\
Dropout rate & 0.05 to 0.40, step 0.05 \\
Learning rate & $10^{-4}$ to $10^{-2}$, logarithmic scale \\
\bottomrule
\end{tabular}
\end{table}

With the hyperparameters fixed, the parameters are estimated by minimizing the
mean squared error between the scaled predictions and the scaled indicators of
the training subset. This loss is a nonconvex function of a large number of
parameters, evaluated on mini-batches of the data, so a stochastic gradient
method is used. The Adam optimizer \cite{kingma2015} is adopted because it
adapts the step of each parameter from running estimates of the first and
second moments of the gradient, which suits gradients of different scales
across the layers and the noise of mini-batch gradients, and because it
requires little tuning beyond the learning rate, which is taken from the first
stage. The training is controlled by the validation subset: the learning rate
is reduced when the validation loss stops decreasing, and the training stops
when the validation mean absolute error does not decrease for a given number of
epochs, with the weights of the best epoch restored, which limits the
overfitting of the training subset. The trained network is evaluated on the
test subset, which is used neither in the training nor in the selection of the
hyperparameters, by the coefficient of determination, the mean absolute error
and the root mean squared error of each output, after the predictions are
brought back to physical units.

% -----------------------------------------------------------------------------
\subsection{Retraining of the neural network}\label{sec:ol}
% -----------------------------------------------------------------------------
Retraining is started by a trigger computed from the plant indicators, uses
data generated at the plant around the operating point in force and updates the
network by fine-tuning.

\begin{figure}[!htb]
\centering
\includegraphics[width=\linewidth]{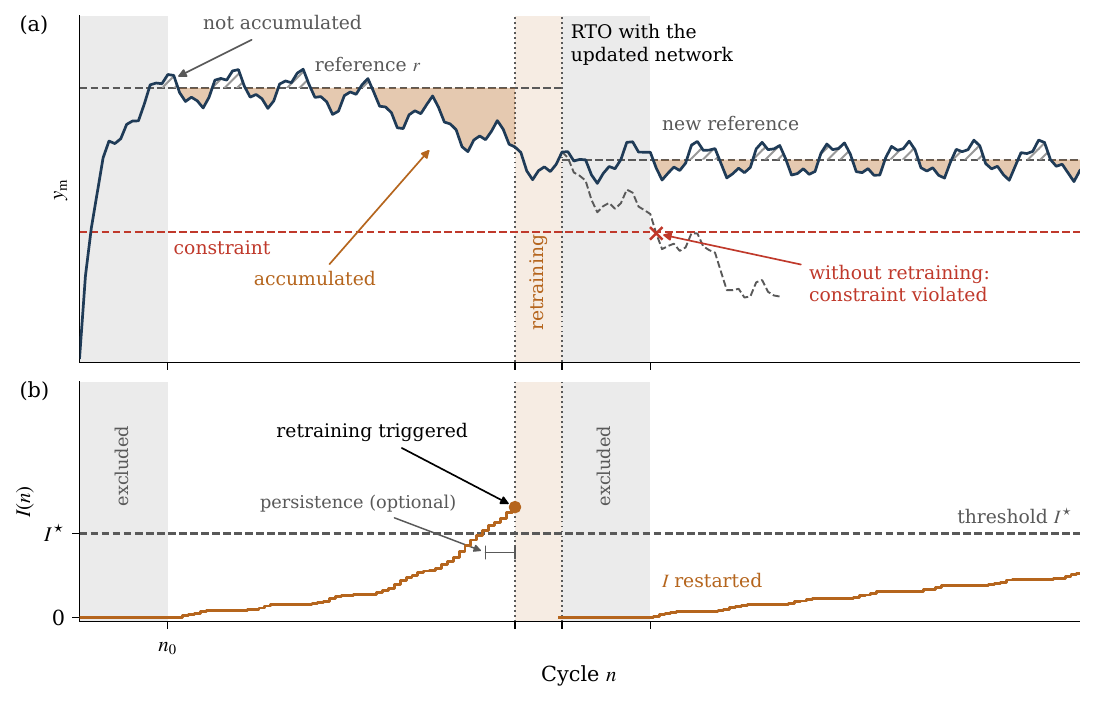}
\caption{Retraining trigger and correction (illustrative values). (a) Monitored
indicator, with the accumulated deviations below the reference and, dashed,
the trajectory without retraining. (b) Accumulated error $I(n)$ and threshold
$I^{\star}$.}
\label{fig:gatilho}
\end{figure}

\label{sec:trigger}Waiting for a constraint violation at CSS to start retraining means that the
plant operates outside specification while the plant settles and the
retraining data are generated, and under a progressive disturbance the
violation may only appear after many cycles. The trigger is therefore based on
the deviation of the plant from the performance achieved after the last RTO
event. A plant indicator $y_{\mathrm{m}}$ is monitored, its value at CSS after
that event is taken as reference $r$, and the relative deviation of the plant
in the unfavourable direction is accumulated cycle by cycle,
\begin{equation}
I(n)=\sum_{i=n_0}^{n}\max\!\left(0,\ \frac{r-y_{\mathrm{m}}(i)}{r}\right),
\label{eq:integral}
\end{equation}
written for an indicator in which a decrease is unfavourable, such as a
purity, a recovery or a productivity; for an indicator in which an increase is
unfavourable, such as the energy per cycle, the sign of the numerator is
reversed. In Eq.~\eqref{eq:integral}, $n_0$ excludes the first cycles of the
monitored phase. Retraining is triggered when $I(n)\ge I^{\star}$, where
$I^{\star}$ is a threshold, and the condition may in addition be required to
hold for a number of consecutive cycles. Only deviations in the unfavourable direction are accumulated, so
that cycles in which the plant performs better than the reference do not offset
the deviation already accrued. The sum depends on both the magnitude and the
duration of the deviation. Fig.~\ref{fig:gatilho} illustrates the procedure
under a progressive disturbance: retraining is triggered when $I(n)$ reaches
$I^{\star}$, the RTO with the updated network moves the plant to a new
operating point before the monitored indicator violates its constraint, and
the monitoring restarts with a new reference.

\label{sec:local}When retraining is triggered, local samples are generated by perturbing the
operating point in force. Each sample is drawn at random, with uniform
distribution, in a box around the current value of each decision variable,
limited to the bounds of Problem~\eqref{eq:rto}, and is run from the current
state until CSS at the plant condition chosen for the data, discussed below. Samples that do
not reach CSS within the maximum number of cycles allowed for each sample are
discarded.

The network is then fine-tuned on the valid samples with the Adam optimizer,
starting from its current weights and keeping the original input and output
scaling, so that the updated network replaces the previous one in the
optimizer without further changes. The hyperparameter search is not repeated.
A low learning rate is used, since the purpose is to shift the network towards
the new condition and not to identify it again, and part of the samples is
reserved for validation and early stopping.

\label{sec:cond}The plant condition at which the local samples are generated
depends on the type of disturbance. After a step, it is the condition that the
RTO will face. Under a progressive disturbance this is
not the case: the operating point chosen after retraining will be applied while
the disturbance continues to evolve. Two choices of the retraining condition
are considered.
\begin{enumerate}[label=(\roman*), leftmargin=2em]
\item Range: the value of the disturbance in each sample is drawn over the
      interval that remains to be traversed, from its value at the trigger to
      its final value.
\item Target: all samples are generated at the final value of the
      disturbance, which the process approaches monotonically.
\end{enumerate}
The target condition assumes that the final value is known.

The samples can be generated in one or two passes. With two passes, the
samples of the first are generated in the box around the operating point in
force, the network is fine-tuned and Problem~\eqref{eq:rto} is solved. Further
samples are then generated in a smaller box around the point proposed by this
optimization, the network in use before retraining is fine-tuned on the valid
samples of both passes and the final optimization is performed, so that the
second pass adds data in the region of the new operating point.

% -----------------------------------------------------------------------------
\subsection{Optimization}\label{sec:rto}
% -----------------------------------------------------------------------------
Each RTO event solves
\begin{equation}
\begin{aligned}
\max_{\mathbf{u}}\quad & J\big(\hat{\mathbf{y}}(\mathbf{u})\big)\\
\text{s.t.}\quad & \hat{y}_j(\mathbf{u})\ \ge\ y_j^{\min}+\beta,
\qquad j\in\mathcal{C},\\
& \mathbf{g}(\mathbf{u})\le \mathbf{0},\qquad
\mathbf{u}^{\mathrm{L}}\le\mathbf{u}\le\mathbf{u}^{\mathrm{U}},
\end{aligned}
\label{eq:rto}
\end{equation}
where $\hat{\mathbf{y}}$ is the surrogate prediction, $J$ is the objective,
$\mathcal{C}$ is the set of constrained indicators, $\beta$ is a back-off
applied to their lower bounds during the optimization, and $\mathbf{g}$
collects constraints on the decision variables. The nominal limits
$y_j^{\min}$ are the ones evaluated at the plant. Upper limits on indicators,
when present, are imposed without back-off.

Problem~\eqref{eq:rto} is solved by a particle swarm optimizer
\cite{kennedy1995} with the feasibility-enhanced constraint handling of
Hasanoglu and Dolen \cite{hasanoglu2018fepso,hasanoglu2018mofepso}, in which a
particle that violates constraints is guided by one of them at a time. Particle swarm optimization has been used in cyclic
adsorption processes, as in the determination of the feasible operating region
of a true moving bed unit \cite{nogueira2019} and the surrogate-assisted
optimization of a vacuum PSA unit for \cotwo{} capture \cite{alkebsi2021}. Each particle $i$ has a position $\mathbf{x}_i$, a
candidate operating point, and a velocity $\mathbf{v}_i$. The swarm is
initialized by Latin hypercube sampling of the bounds, with one
particle placed at the operating point in force, so that the current point is
always among the candidates. During the initialization the limits of the
constrained indicators are relaxed, so that feasible particles are available
to guide the swarm, and the actual limits are imposed before the first
iteration. At each iteration, a particle that
satisfies all constraints moves according to
\begin{equation}
\mathbf{v}_i\leftarrow w\,\mathbf{v}_i
+c_1\,\mathbf{r}_1\odot(\mathbf{g}^{\mathrm{best}}-\mathbf{x}_i)
+c_2\,\mathbf{r}_2\odot(\mathbf{p}_i-\mathbf{x}_i),
\qquad
\mathbf{x}_i\leftarrow\mathbf{x}_i+\mathbf{v}_i,
\label{eq:pso}
\end{equation}
where $\mathbf{g}^{\mathrm{best}}$ is the best feasible position found by the
swarm, $\mathbf{p}_i$ the best feasible position found by particle $i$,
$\mathbf{r}_1$ and $\mathbf{r}_2$ are vectors of random numbers uniformly
distributed in $[0,1]$, $c_1$ and $c_2$ are the acceleration coefficients, and
the inertia weight $w$ decreases linearly along the iterations. A particle that violates
constraints selects one of them, giving priority to the constraints it has
selected least often and, among these, to the one violated by the largest
number of particles. It then moves towards a randomly chosen particle
$\mathbf{x}_k$ that satisfies the selected constraint,
$\mathbf{v}_i\leftarrow w\,\mathbf{v}_i+c_1\,\mathbf{r}_1\odot(\mathbf{x}_k-\mathbf{x}_i)$,
and only the decision variables that affect the selected constraint are
updated. This dependence is determined once, after the initialization, by
perturbing each variable of each particle by a fraction of its range and
recording which constraints change state. A position that leaves the bounds is
brought back along the direction of the move, and the objective is evaluated
only at feasible positions. The optimum is the best feasible point among the
best positions stored and the current positions of the swarm.

An RTO event is executed when the plant is at CSS. The optimum of
Problem~\eqref{eq:rto} is implemented in a single move, without trust region or
ramping of the decision variables, and the plant is then integrated from its
current state until CSS is reached and for a fixed number of further cycles. The surrogate
prediction at the optimum is stored together with the plant indicators at the
new CSS, which gives a direct measure of the model error at each event. The
plant integration is continuous across events, so each event starts from the
physical state left by the previous one.\label{sec:exec}

\FloatBarrier
% =============================================================================
\section{Case study: \chfour/\cotwo{} separation by PSA}\label{sec:case}
% =============================================================================
The methodology is applied to the separation of \chfour{} and \cotwo{} by
PSA. In biogas upgrading, the raw gas must be brought to a \chfour{} content
that meets the specification of biomethane, and PSA is one of the established
technologies for this separation \cite{sun2015,angelidaki2018}, studied with
equilibrium and kinetic adsorbents and with different cycle configurations
\cite{grande2007a,augelletti2017}. This
section describes the process, the virtual plant and the disturbances, and the
performance indicators.

\subsection{Process and virtual plant}\label{sec:plant}
The unit separates a \chfour/\cotwo{} feed with a six-step cycle of the
Skarstrom type with a heavy reflux (rinse) step, listed in
Table~\ref{tab:steps}. The feed is
supplied at $F=5.0$\,mol\,s$^{-1}$, 303.15\,K and a \cotwo{} mole fraction
$y_{\cotwo}=0.40$, except where a disturbance is applied, as described at the
end of this section. The high and low pressures are 5.0 and 0.1\,bar.
\chfour{} is recovered as the light product during feed and rinse, and \cotwo{}
as the heavy product during blowdown and purge.

The cycle is defined by five decision variables: the feed step duration
$t_{\mathrm{F}}$, the ratio $r_{\mathrm{R}}$ between rinse and feed durations,
and three split fractions. The rinse duration is
$t_{\mathrm{R}}=r_{\mathrm{R}}t_{\mathrm{F}}$ and the blowdown duration is
$t_{\mathrm{B}}=t_{\mathrm{F}}-t_{\mathrm{R}}$. The purge lasts
$t_{\mathrm{F}}$, and the pressurization, which ends when the high pressure is
reached, is
followed by an idle step that completes a period of
$t_{\mathrm{F}}$. The cycle therefore lasts $4\,t_{\mathrm{F}}$, of which the
feed step occupies one quarter at any operating point. The rinse step is fed with a fraction $s_{\mathrm{R}}$
of the heavy product leaving blowdown and purge. A fraction $s_{\mathrm{L}}$ of
the light product of feed and rinse is returned to the column as light reflux
and is divided between purge, which receives $s_{\mathrm{L}}s_{\mathrm{P}}$,
and pressurization, which receives $s_{\mathrm{L}}(1-s_{\mathrm{P}})$. The
vector of decision variables is therefore
$\mathbf{u}=[t_{\mathrm{F}},\,r_{\mathrm{R}},\,s_{\mathrm{R}},\,
s_{\mathrm{L}},\,s_{\mathrm{P}}]$.

\begin{table}[htbp]
\centering
\caption{Steps of the PSA cycle.}
\label{tab:steps}
\small
\begin{tabular}{clll}
\toprule
Step & Name & Flow direction relative to the feed & Inlet stream \\
\midrule
1 & Feed           & co-current      & fresh feed \\
2 & Rinse          & co-current      & fraction $s_{\mathrm{R}}$ of steps 3 and 4 outlet \\
3 & Blowdown       & counter-current & none (closed inlet) \\
4 & Purge          & counter-current & fraction $s_{\mathrm{L}}s_{\mathrm{P}}$ of steps 1 and 2 outlet \\
5 & Pressurization & counter-current & fraction $s_{\mathrm{L}}(1-s_{\mathrm{P}})$ of steps 1 and 2 outlet \\
6 & Idle           & none     & none \\
\bottomrule
\end{tabular}
\end{table}

The virtual plant is a one-dimensional, non-isothermal model of a packed
column with linear driving force mass transfer, Langmuir equilibrium and the
Ergun equation for the pressure drop, integrated cycle by cycle. It follows
the assumptions of the PSA model of our previous work \cite{rebello2024} and
is given in
Appendix~\ref{sec:model}. The same model generates the data used to build and
retrain the surrogate.

Two disturbances are considered, each in its own scenario\label{sec:dist}. Neither is an
input of the surrogate model.

\textit{Feed composition step.} The \cotwo{} mole fraction of the feed changes
from 0.40 to 0.50 at constant feed flow rate and constant operating point.

\textit{Sorbent ageing.} The loss of capacity is represented by the activity
$a$ in Eq.~\eqref{eq:iso} (Appendix~\ref{sec:model}), which scales the
saturation capacity of both components and leaves the affinity constants
unchanged. This corresponds to a loss of accessible sites with the selectivity
preserved. The activity decreases linearly with the cycle number, by 0.003125
per cycle, from $a=1.00$ to $a=0.50$, which is reached after 160 cycles.

\subsection{Performance indicators and cyclic steady state}\label{sec:kpi}
The indicators are computed from the amounts of each component entering and
leaving the column in each step of a cycle: $N^{\mathrm{in}}_{i,k}$ and
$N^{\mathrm{out}}_{i,k}$ are the amounts of component $i$ entering and leaving
the column in step $k$, and $N^{\mathrm{in}}_{k}$ and $N^{\mathrm{out}}_{k}$
the total amounts. \chfour{} is collected in steps 1 and 2, and part of it
returns to the column in steps 4 and 5; \cotwo{} is collected in steps 3 and
4, and part of it returns in step 2. The net amounts produced in a cycle are
therefore
\begin{equation}
N^{\mathrm{net}}_{\chfour}=N^{\mathrm{out}}_{\chfour,1}+N^{\mathrm{out}}_{\chfour,2}
-N^{\mathrm{in}}_{\chfour,4}-N^{\mathrm{in}}_{\chfour,5},
\qquad
N^{\mathrm{net}}_{\cotwo}=N^{\mathrm{out}}_{\cotwo,3}+N^{\mathrm{out}}_{\cotwo,4}
-N^{\mathrm{in}}_{\cotwo,2}.
\label{eq:net}
\end{equation}
The purities of the two products are
\begin{equation}
P_{\chfour}=\frac{N^{\mathrm{out}}_{\chfour,1}+N^{\mathrm{out}}_{\chfour,2}}
{N^{\mathrm{out}}_{1}+N^{\mathrm{out}}_{2}},
\qquad
P_{\cotwo}=\frac{N^{\mathrm{out}}_{\cotwo,3}+N^{\mathrm{out}}_{\cotwo,4}}
{N^{\mathrm{out}}_{3}+N^{\mathrm{out}}_{4}},
\label{eq:kpi}
\end{equation}
and the recovery and the productivity of each component $i$ are
\begin{equation}
R_{i}=\frac{N^{\mathrm{net}}_{i}}{N^{\mathrm{in}}_{i,1}},
\qquad
Pr_{i}=\frac{N^{\mathrm{net}}_{i}}{t_{\mathrm{c}}\,m_{\mathrm{s}}},
\label{eq:recprod}
\end{equation}
where $t_{\mathrm{c}}=4\,t_{\mathrm{F}}$ is the duration of the cycle and
$m_{\mathrm{s}}$ the sorbent mass. The energy per cycle is the isentropic work
of compressing the gas admitted in the feed and pressurization steps from the
low pressure $P_{\mathrm{L}}$ to the high pressure $P_{\mathrm{H}}$,
\begin{equation}
E=\frac{N^{\mathrm{in}}_{1}+N^{\mathrm{in}}_{5}}{\eta}\,
\frac{\gamma}{\gamma-1}\,R\,T_{\mathrm{in}}
\left[\left(\frac{P_{\mathrm{H}}}{P_{\mathrm{L}}}\right)^{(\gamma-1)/\gamma}-1\right],
\label{eq:energia}
\end{equation}
where $\eta=0.72$ is the efficiency of the compressor, $T_{\mathrm{in}}$ the
feed temperature and $\gamma$ the ratio of heat capacities of the feed
mixture. Since $E$ depends on the duration of the cycle, the specific energy,
\begin{equation}
E_{\mathrm{s}}=\frac{E}{N^{\mathrm{net}}_{\chfour}},
\label{eq:es}
\end{equation}
is used to compare operating points with different cycle durations.

CSS is declared when, over a window of five consecutive cycles, the relative
amplitude of each purity and recovery is below 0.25\,\%, that of the
productivities and of the energy is below 1\,\%, and the overall mass balance
of each component closes within 0.5\,\% in every cycle of the window.

\FloatBarrier
% =============================================================================
\section{Results and discussion}\label{sec:res}
% =============================================================================
The neural network has six inputs, the feed flow rate and the five decision
variables, and seven outputs: the purity and recovery of both components, the
productivity of both components and the energy per cycle. The data were
generated as in Section~\ref{sec:aq} at $y_{\cotwo}=0.40$ and $a=1$, from
11\,000 samples of optimized Latin hypercube designs with
$t_{\mathrm{F}}\in[10,800]$\,s, $r_{\mathrm{R}}\in[0.05,0.95]$ and split
fractions in $[0.01,0.99]$. Each sample was integrated for at most 150 cycles,
which gave 8613 points at CSS, divided into 6029 points for training and 1292
each for validation and test. The feed flow rate is constant in the data and
was kept at 5.0\,mol\,s$^{-1}$ in all scenarios. Neither the feed composition
nor the sorbent activity is an input of the network, so two plant states that
differ only in $y_{\cotwo}$ or in $a$ correspond to the same input and the
original network cannot distinguish them.

The network was identified as described in Section~\ref{sec:surr}. The
Hyperband search, with
a reduction factor $\eta=3$ and a maximum of 100 epochs per configuration,
evaluated 254 configurations, and the selected hyperparameters are listed in
Table~\ref{tab:hpbest}: four hidden layers with ReLU activation, without batch
normalization or dropout, followed by a linear output layer. The selected
network was trained in mini-batches of 32 points for at most 1000 epochs; the
learning rate was halved when the validation loss did not decrease for 15
epochs, down to $10^{-6}$, and the training stopped when the validation mean
absolute error did not decrease for 40 epochs. On the test subset, the coefficient of determination
ranges from 0.997, for the \chfour{} purity, to 0.9998, for the energy per
cycle, across the seven outputs. This network is the surrogate used at the
start of the three scenarios reported below, and it is the network that is
fine-tuned when retraining is triggered in scenarios 2 and 3.

\begin{table}[htbp]
\centering
\caption{Hyperparameters selected by the Hyperband search.}
\label{tab:hpbest}
\small
\begin{tabular}{ll}
\toprule
Hyperparameter & Selected value \\
\midrule
Hidden layers & 4 \\
Neurons per hidden layer & 384, 512, 384, 512 \\
Activation function & ReLU \\
Batch normalization & no \\
Dropout & no \\
Learning rate & $4.9\times10^{-4}$ \\
\bottomrule
\end{tabular}
\end{table}

% -----------------------------------------------------------------------------
\subsection{Scenario 1: sequential RTO under changing objectives}\label{sec:caso1}
% -----------------------------------------------------------------------------
Scenario 1 applies the RTO without disturbance, to establish the agreement
between surrogate and plant when both describe the same process. The plant
starts at the operating point
$\mathbf{u}_0=[200\ \mathrm{s},\,0.40,\,0.30,\,0.50,\,0.20]$, at which it
reaches CSS after 14 cycles with a \chfour{} purity of 87.35\,\%. Five RTO
events were then executed in sequence, with the following objectives and
nominal constraints:
\begin{equation}
\begin{alignedat}{3}
&\text{RTO 1:}\quad && \max_{\mathbf{u}}\ Pr_{\chfour} &&\quad\text{s.t.}\quad
  P_{\chfour}\ge 90\,\%,\ R_{\chfour}\ge 70\,\%,\\
&\text{RTO 2:}\quad && \max_{\mathbf{u}}\ R_{\chfour} &&\quad\text{s.t.}\quad
  P_{\chfour}\ge 95\,\%,\\
&\text{RTO 3:}\quad && \max_{\mathbf{u}}\ R_{\cotwo} &&\quad\text{s.t.}\quad
  P_{\cotwo}\ge 85\,\%,\\
&\text{RTO 4:}\quad && \max_{\mathbf{u}}\ Pr_{\chfour} &&\quad\text{s.t.}\quad
  P_{\chfour}\ge 80\,\%,\ R_{\chfour}\ge 70\,\%,\ E\le 30\ \mathrm{MJ\,cycle^{-1}},\\
&\text{RTO 5:}\quad && \min_{\mathbf{u}}\ E &&\quad\text{s.t.}\quad
  P_{\chfour}\ge 80\,\%,\ R_{\chfour}\ge 70\,\%,\ P_{\cotwo}\ge 50\,\%,\
  R_{\cotwo}\ge 50\,\%.
\end{alignedat}
\label{eq:prob1}
\end{equation}
Each problem is solved in the form of Problem~\eqref{eq:rto}, with
$\beta=2$\pp{} The bounds on the decision variables are
$t_{\mathrm{F}}\in[60,800]$\,s, $r_{\mathrm{R}}\in[0.05,0.80]$,
$s_{\mathrm{R}},s_{\mathrm{L}}\in[0.01,0.99]$ and
$s_{\mathrm{P}}\in[0.05,0.95]$, and the constraints $\mathbf{g}$ are
$t_{\mathrm{B}}\ge 60$\,s, $t_{\mathrm{R}}\ge 10$\,s,
$s_{\mathrm{L}}s_{\mathrm{P}}\ge 0.05$ and
$s_{\mathrm{L}}(1-s_{\mathrm{P}})\ge 0.02$. These constraints keep the cycle
within the region where the rigorous model converges to a stable CSS, since
small purge or pressurization flows produced alternating cycle-to-cycle
oscillations of the solution. The particle swarm has 100 particles and runs
for 200 iterations, with $c_1=c_2=2$ and $w$ decreasing from 0.9 to 0.4; the
limits of the constrained indicators are relaxed by 10\pp{} during the
initialization, and the dependence of the constraints on the decision
variables is determined with perturbations of $\pm30\,\%$ of the range of each
variable. After each RTO event the plant is integrated until CSS and for a
further 20 cycles. The same settings are used in scenarios 2 and 3.

The sequence covers \chfour{} production at two purity specifications,
\cotwo{} recovery, \chfour{} production under an energy ceiling and minimum
energy consumption at relaxed specifications, so that consecutive optima lie
in different regions of the operating domain.

\begin{figure}[htbp]
\centering
\includegraphics[width=0.95\linewidth,height=0.86\textheight,keepaspectratio]{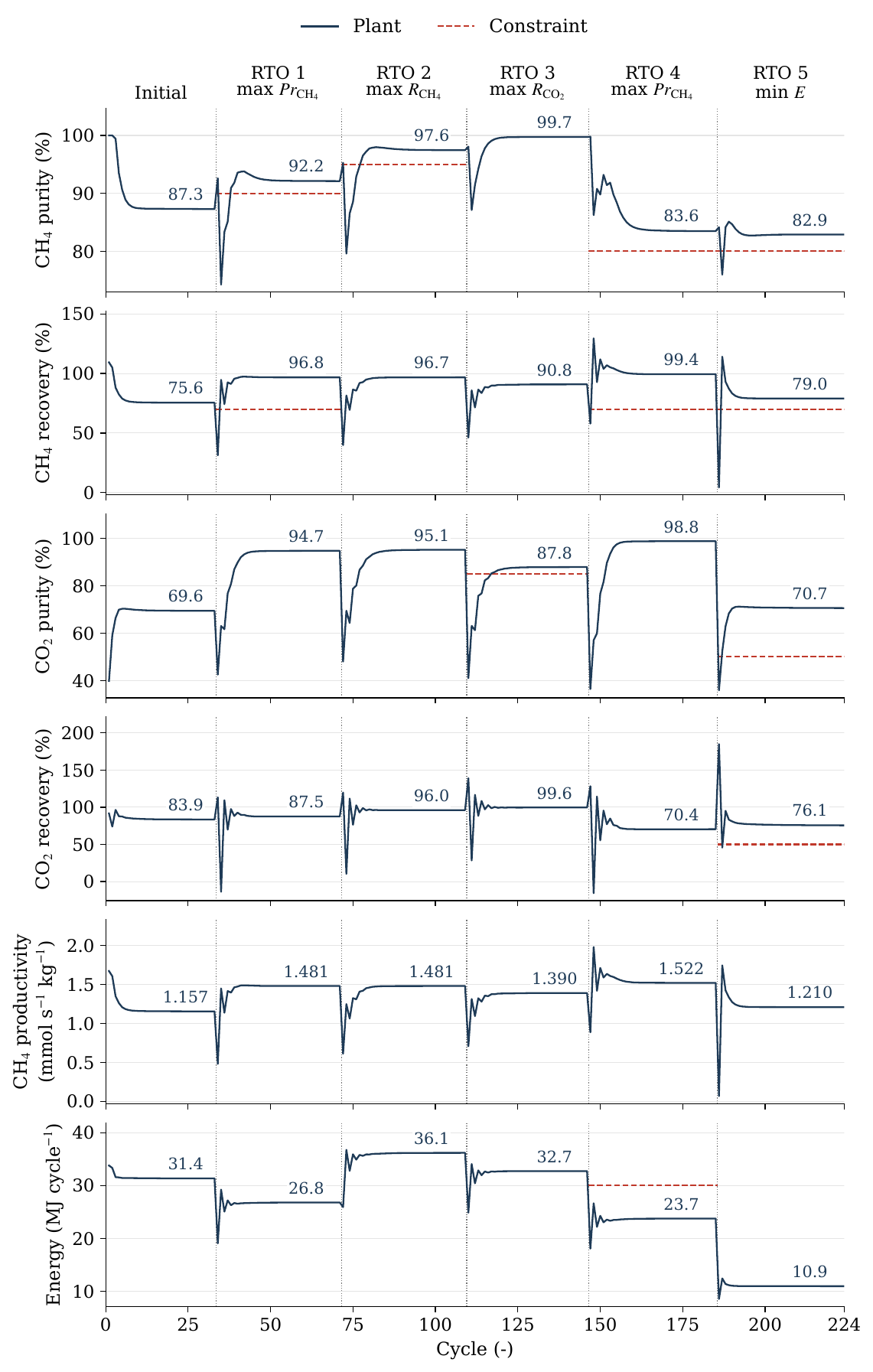}
\caption{Scenario 1: plant indicators over five consecutive RTO events,
with the objective of each event above the phase. The numbers are the values
at CSS in each phase, and
the dashed lines are the constraints of Eq.~\eqref{eq:prob1}.
Recoveries above 100\,\% or below zero occur only in the first cycles after a
move, while the inventory of the bed changes.}
\label{fig:caso1}
\end{figure}

Fig.~\ref{fig:caso1} shows the cycle-by-cycle trajectory of the six
indicators at the plant, with the objective of each RTO event, its
constraints and the value reached at CSS in each phase. Table~\ref{tab:kpi1}
lists the plant indicators at CSS and the deviation from the surrogate
prediction, and Table~\ref{tab:op1} the operating point of each phase. The
plant reached CSS between 18 and 20 cycles after each move, and all nominal
constraints were satisfied at CSS in the five phases.

\begin{table}[htbp]
\centering
\caption{Scenario 1: plant indicators at CSS and largest absolute deviation
between surrogate prediction and plant over the four purities and recoveries.}
\label{tab:kpi1}
\small
\begin{tabular}{lcccccccc}
\toprule
Phase & $P_{\chfour}$ & $R_{\chfour}$ & $P_{\cotwo}$ & $R_{\cotwo}$ &
$Pr_{\chfour}$ & $E$ & $E_{\mathrm{s}}$ & Deviation \\
 & (\%) & (\%) & (\%) & (\%) & (mmol\,s$^{-1}$\,kg$^{-1}$) &
(MJ\,cycle$^{-1}$) & (kJ\,mol$^{-1}$) & (p.p.) \\
\midrule
Initial & 87.35 & 75.58 & 69.57 & 83.86 & 1.157 & 31.35 & 69.1 & n.a. \\
RTO 1   & 92.20 & 96.77 & 94.73 & 87.52 & 1.481 & 26.77 & 66.4 & 0.69 \\
RTO 2   & 97.57 & 96.73 & 95.08 & 96.05 & 1.481 & 36.14 & 74.8 & 1.36 \\
RTO 3   & 99.73 & 90.82 & 87.80 & 99.58 & 1.390 & 32.72 & 81.1 & 0.42 \\
RTO 4   & 83.61 & 99.44 & 98.79 & 70.41 & 1.522 & 23.73 & 69.6 & 0.56 \\
RTO 5   & 82.88 & 79.04 & 70.72 & 76.12 & 1.210 & 10.95 & 61.0 & 1.70 \\
\bottomrule
\end{tabular}
\end{table}

\begin{table}[htbp]
\centering
\caption{Scenario 1: operating point of each phase.}
\label{tab:op1}
\small
\begin{tabular}{lccccc}
\toprule
Phase & $t_{\mathrm{F}}$ (s) & $r_{\mathrm{R}}$ & $s_{\mathrm{R}}$ &
$s_{\mathrm{L}}$ & $s_{\mathrm{P}}$ \\
\midrule
Initial & 200.0 & 0.400 & 0.300 & 0.500 & 0.200 \\
RTO 1   & 138.8 & 0.328 & 0.705 & 0.577 & 0.138 \\
RTO 2   & 166.4 & 0.291 & 0.708 & 0.639 & 0.086 \\
RTO 3   & 148.0 & 0.151 & 0.692 & 0.671 & 0.107 \\
RTO 4   & 114.4 & 0.277 & 0.796 & 0.573 & 0.089 \\
RTO 5   &  75.7 & 0.166 & 0.541 & 0.500 & 0.466 \\
\bottomrule
\end{tabular}
\end{table}

Two relations of this process help to read the results. First, at CSS the
overall balance links the four purities and recoveries: the \cotwo{} that is
not recovered in the heavy product leaves with the light product and lowers
the \chfour{} purity, and the \chfour{} that is not recovered leaves with the
heavy product and lowers the \cotwo{} purity. Second, the feed flow rate is
fixed and the feed step occupies one quarter of the cycle
(Section~\ref{sec:plant}), so the amount of gas processed per unit time is the
same at all operating points. For a given feed composition, the \chfour{} productivity is then proportional
to the \chfour{} recovery, and the energy per cycle, which is the work of
compressing the gas admitted in the feed and pressurization steps, increases
with $t_{\mathrm{F}}$; the specific energy $E_{\mathrm{s}}$ removes this effect
of the cycle duration.

The first event increased the \chfour{} productivity by 28\,\% relative to the
initial point and raised the purity from 87.4 to 92.2\,\%, above the 90\,\%
specification that the initial point did not meet. The main changes in the
operating point were the larger fraction of the heavy product recycled in the
rinse step ($s_{\mathrm{R}}$ from 0.30 to 0.71) and the shorter feed step (139
against 200\,s). The rinse displaces the \chfour{} held in the bed towards the
light product before the blowdown \cite{ruthven1994}, so less \chfour{} leaves
with the heavy product: the \chfour{} recovery increased from 75.6 to
96.8\,\% and the \cotwo{} purity from 69.6 to 94.7\,\%. The shorter feed step
admits less \cotwo{} per cycle, which is consistent with the higher \chfour{}
purity.

With the purity specification raised to 95\,\% and the recovery as objective
(RTO 2), the purity reached 97.6\,\% with productivity and recovery practically
unchanged. The light reflux increased ($s_{\mathrm{L}}$ from 0.58 to 0.64), with
a larger share directed to the pressurization, which fills the column with
\chfour{}-rich gas from the product end. The higher purity was obtained at the
expense of energy: more gas is recompressed in the pressurization and the feed
step is longer (166\,s), so the energy per cycle increased by 35\,\% and the
specific energy by 13\,\%. Directing the objective to \cotwo{} recovery (RTO 3)
gave a recovery of 99.6\,\% with a \cotwo{} purity of 87.8\,\%, close to its
85\,\% limit. Almost no \cotwo{} reaches the light product, whose purity is
99.7\,\%, but more \chfour{} leaves with the heavy product: the \chfour{}
recovery decreased to 90.8\,\%, the productivity is 6\,\% lower than in RTO 2,
and the specific energy is the highest of the sequence, with the largest
light reflux ($s_{\mathrm{L}}=0.67$) and less \chfour{} recovered.

The lower purity floor of RTO 4 moved the operating point in the opposite
direction. More \cotwo{} is allowed into the light product (\cotwo{} recovery
of 70.4\,\%), the heavy product is almost pure \cotwo{} (98.8\,\%) and almost
no \chfour{} is lost with it (recovery of 99.4\,\%), which gives the highest
productivity of the sequence, 2.8\,\% above RTO 1, at a purity 8.6\pp{} lower.
At this optimum the energy ceiling was not active, since the plant consumed
23.7\,MJ per cycle. The minimum-energy event (RTO 5) reduced the energy per
cycle to 10.9\,MJ, 65\,\% below the initial point, with all relaxed
specifications satisfied. Most of this reduction reflects the shorter cycle,
since the feed step went to 76\,s, close to its lower bound, and the fraction
of the light product used in the pressurization is the smallest of the
sequence; per mole of \chfour{} produced, the energy
decreased from 69.1 to 61.0\,kJ\,mol$^{-1}$, 12\,\% below the initial point.

The deviation between surrogate and plant did not exceed 1.7\pp{} in any
purity or recovery, and the predicted energy was within 1.5\,\% of the plant
value in the first four events. For the constraints that were active at the
optimum, the plant value was above the prediction in the four events in which
such a constraint existed: by 0.10\pp{} for the \chfour{} purity in RTO 1,
0.48\pp{} in RTO 2, 0.19\pp{} for the \cotwo{} purity in RTO 3 and 0.85\pp{}
for the \chfour{} purity in RTO 5. The largest relative discrepancy occurred in
the energy of RTO 5, in which the plant consumed 4.4\,\% more energy than
predicted, without effect on the constraints of that event. An optimization
with these settings takes between 83 and 98\,s on a virtual machine with two
cores of an Intel Xeon processor at 2.1\,GHz, without GPU, which is less than
the duration of one cycle at any of the operating points of
Table~\ref{tab:op1} (303 to 800\,s).

Each optimum was applied in a single move, and the transients that followed
were large. After RTO 1 the \chfour{} purity fell to 74\,\% and remained below
90\,\% for three cycles, and after RTO 3 the \cotwo{} purity remained below
85\,\% for seven cycles. The constraints of Problem~\eqref{eq:rto} refer to CSS
values, and these excursions are outside the scope of the steady-state RTO
layer. Limiting them would require a transition strategy, such as a gradual
change of the decision variables or a dynamic control layer, which was not
included in order to observe the response to the RTO decisions directly.

Without disturbance, the surrogate-based RTO returned feasible optima for five
objectives with different active constraints, with deviations between
surrogate and plant of at most 1.7\pp{} Scenarios 2 and 3 start from this
condition, so the loss of feasibility observed there can be attributed to the
disturbance.

\FloatBarrier
% -----------------------------------------------------------------------------
\subsection{Scenario 2: online learning after a step in feed composition}\label{sec:caso2}
\begin{figure}[htbp]
\centering
\includegraphics[width=0.95\linewidth,height=0.86\textheight,keepaspectratio]{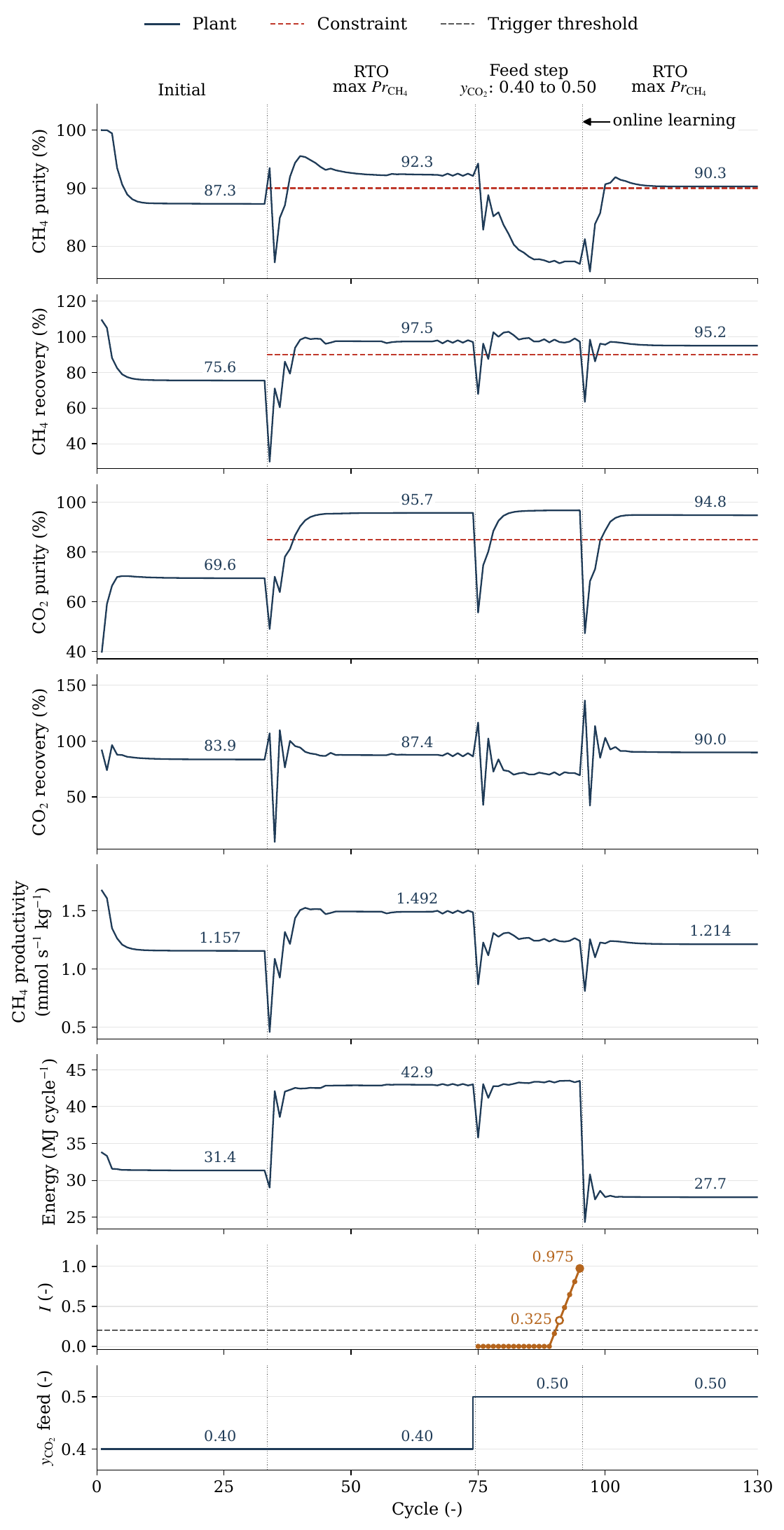}
\caption{Scenario 2: plant indicators, accumulated error $I$ and \cotwo{}
feed fraction in phases I to IV, with the objective of each RTO event above the phase. The numbers are
the values at CSS in each phase, and the dashed lines are the constraints and
the trigger threshold. The open circle marks the cycle in which $I$ crossed the
threshold and the filled circle the cycle in which retraining was triggered.}
\label{fig:caso2}
\end{figure}

% -----------------------------------------------------------------------------
Scenario 2 introduces a disturbance that is not represented in the surrogate
inputs, a step in feed composition. It has four phases
(Fig.~\ref{fig:caso2}): (I) start-up at the initial point $\mathbf{u}_0$ of
scenario 1; (II) one RTO event at the nominal feed composition, whose CSS
purity defines the reference $r$ of Eq.~\eqref{eq:integral}; (III) a step in
the \cotwo{} feed fraction from 0.40 to 0.50, with the operating point of
phase II kept and the trigger monitored; and (IV) local data generation,
fine-tuning and a new RTO event. The RTO events of phases II and IV solve
\begin{equation}
\max_{\mathbf{u}}\ Pr_{\chfour}\quad\text{s.t.}\quad
P_{\chfour}\ge 90\,\%,\ R_{\chfour}\ge 90\,\%,\ P_{\cotwo}\ge 85\,\%,
\label{eq:prob2}
\end{equation}
in the form of Problem~\eqref{eq:rto}, with the settings of scenario 1. In the
event of phase IV, the rinse recycle intensity was also limited to
$s_{\mathrm{R}}s_{\mathrm{L}}\le 0.45$, for the reason given below. The trigger
monitors the \chfour{} purity, with $I^{\star}=0.20$, the first 15 cycles after
the step excluded from Eq.~\eqref{eq:integral} and a persistence of five
consecutive cycles.

\begin{table}[htbp]
\centering
\caption{Scenario 2: surrogate prediction and plant value. Phases II and IV
at CSS; phase III at the cycle in which retraining was triggered.}
\label{tab:caso2}
\small
\begin{tabular}{llccccc}
\toprule
Phase & Surrogate & $y_{\cotwo}$ & \multicolumn{2}{c}{$P_{\chfour}$ (\%)} &
\multicolumn{2}{c}{$Pr_{\chfour}$ (mmol\,s$^{-1}$\,kg$^{-1}$)} \\
\cmidrule(lr){4-5}\cmidrule(lr){6-7}
 & & & Predicted & Plant & Predicted & Plant \\
\midrule
II  & original   & 0.40 & 92.31 & 92.27 & 1.485 & 1.492 \\
III & original   & 0.50 & 92.31 & 76.95 & 1.485 & 1.241 \\
IV  & fine-tuned & 0.50 & 92.17 & 90.31 & 1.219 & 1.214 \\
\bottomrule
\end{tabular}
\end{table}

At the nominal composition the RTO returned a point predicted at 92.31\,\%
purity, and the plant reached 92.27\,\% (Table~\ref{tab:caso2}), which was
adopted as the reference. At this operating point the indicators alternate
from cycle to cycle (Fig.~\ref{fig:caso2}), with an amplitude of 0.41\pp{} in
the \chfour{} purity and 3.0\pp{} in the \cotwo{} recovery over the last ten
cycles of the phase. The rinse step is fed with the heavy product of the
previous cycle (Appendix~\ref{sec:model}), which couples consecutive cycles,
and with a large recycle, here $s_{\mathrm{R}}s_{\mathrm{L}}=0.50$, the
indicators alternate between two values instead of settling. For this reason
the rinse recycle intensity was limited to
$s_{\mathrm{R}}s_{\mathrm{L}}\le 0.45$ in the RTO event of phase IV, at whose
operating point the amplitude is below 0.05\pp, and in the RTO events of
scenario 3. The operating points of scenario 1, obtained without this limit,
have $s_{\mathrm{R}}s_{\mathrm{L}}\le 0.46$ and do not show the alternation. Since $y_{\cotwo}$ is not a surrogate input and the
operating point did not change in phase III, the prediction in that phase is
identical to that of phase II, and the difference between prediction and plant
measures the mismatch introduced by the step. The \chfour{} purity fell below
90\,\% two cycles after the step and approached 77\,\%. At constant operating
point, the step raises the amount of \cotwo{} admitted in the feed step by
25\,\%, from 383 to 479\,mol per cycle, while the capacity of the bed is
unchanged, so the bed can no longer retain all of it and \cotwo{} breaks
through into the light product. The indicators at the trigger are consistent
with this: the \cotwo{} recovery, which is not constrained, had decreased from
87.4 to 69.4\,\%, that is, the additional \cotwo{} left with the light product,
whereas the \chfour{} recovery and the \cotwo{} purity differed by less than
1.1\pp{} from phase II. The \chfour{} productivity had decreased by 17\,\%,
which is the reduction of the \chfour{} mole fraction of the feed, from 0.60 to
0.50, since the recovery was practically unchanged.

The accumulated error crossed the threshold in the second cycle that was
counted, 17 cycles after the step, with $I=0.325$. Since the deviation
persisted, $I$ remained above the threshold in the following cycles, and
retraining was triggered 21 cycles after the step, once the condition had held
for five cycles, with $I=0.975$ and a purity of 76.95\,\%, 15.3\pp{}
below the reference.

A new optimization with the original surrogate would return the same
operating point, because none of its inputs changed. Forty-five samples were
generated at $y_{\cotwo}=0.50$ in a single pass, drawn in a box of
$\pm40\,\%$ around the operating point of phase II and integrated for at most
40 cycles each, starting from the state of the column at the trigger; 43 of
them reached CSS. The network was fine-tuned on these samples with a learning
rate of $2\times10^{-4}$, halved when the validation loss did not decrease for
25 epochs, a batch size of 8 and 25\,\% of the samples for validation; the
fine-tuning stops when the validation loss does not decrease for 60 epochs or
after 400 epochs, which were completed in this case. Table~\ref{tab:mae} gives the mean absolute error of the network on the
validation samples before and after fine-tuning. Before it, the largest errors
were in the \chfour{} purity (9.3\pp) and in the \cotwo{} recovery (8.8\pp),
the indicators that reflect the breakthrough of \cotwo, and after it the errors
of all purities and recoveries were at most 1.2\pp{} The error in the energy
changed little, from 0.43 to 0.33\,MJ per cycle, about 1\,\% of its value,
since the amount of gas compressed per cycle depends little on the feed
composition.

\begin{table}[htbp]
\centering
\caption{Mean absolute error of the network on the validation samples, before
(original) and after fine-tuning; $n$ is the number of validation samples.}
\label{tab:mae}
\small
\begin{tabular}{lcccccc}
\toprule
 & \multicolumn{2}{c}{Scenario 2 ($n=11$)} &
\multicolumn{2}{c}{Scenario 3, range ($n=10$)} &
\multicolumn{2}{c}{Scenario 3, target ($n=9$)} \\
\cmidrule(lr){2-3}\cmidrule(lr){4-5}\cmidrule(lr){6-7}
Output & Original & Fine-tuned & Original & Fine-tuned & Original & Fine-tuned \\
\midrule
$P_{\chfour}$ (p.p.) & 9.34 & 0.86 & 1.72 & 0.57 & 2.05 & 0.51 \\
$R_{\chfour}$ (p.p.) & 4.31 & 0.59 & 1.18 & 0.80 & 1.81 & 0.21 \\
$P_{\cotwo}$ (p.p.)  & 7.06 & 1.15 & 2.40 & 0.98 & 3.20 & 0.35 \\
$R_{\cotwo}$ (p.p.)  & 8.82 & 0.87 & 3.15 & 1.11 & 3.59 & 0.92 \\
$Pr_{\chfour}$ (mmol\,s$^{-1}$\,kg$^{-1}$) & 0.165 & 0.009 & 0.019 & 0.011 & 0.027 & 0.002 \\
$Pr_{\cotwo}$ (mmol\,s$^{-1}$\,kg$^{-1}$)  & 0.100 & 0.016 & 0.032 & 0.011 & 0.037 & 0.008 \\
$E$ (MJ\,cycle$^{-1}$) & 0.43 & 0.33 & 0.53 & 0.28 & 0.25 & 0.23 \\
\bottomrule
\end{tabular}
\end{table}

The RTO with the fine-tuned network returned a point
predicted at 92.17\,\% purity, and the plant reached 90.31\,\% at CSS, 16 cycles
after the move, restoring feasibility. The deviation between prediction and
plant in the four purities and recoveries was at most 1.9\pp, for the
\chfour{} purity, with the plant below the prediction in all of them. During
the transient that followed the move, the purity remained below 90\,\% for
four cycles.

The new operating point has a feed step 31\,\% shorter (132 against 191\,s),
which brings the \cotwo{} admitted per cycle to 330\,mol, below the 383\,mol of
phase II, and the \cotwo{} recovery returned to 90.0\,\%. Because the cycle is
shorter in the same proportion, the feed flow rate processed is unchanged. The
productivity, 1.214\,mmol\,s$^{-1}$\,kg$^{-1}$, is 19\,\% lower than in phase
II; 17\,\% corresponds to the lower \chfour{} content of the feed and the
remainder to the lower recovery (95.2 against 97.5\,\%). The energy per cycle
decreased to 27.7\,MJ with the shorter cycle, but the specific energy increased
from 76.5 to 88.2\,kJ\,mol$^{-1}$, since a similar compression work per mole of
feed now yields less \chfour{}.

With the operating point of phase II kept, the light product would remain
13.1\pp{} below the 90\,\% specification for as long as the new feed
composition lasted; 20 of the 21 cycles monitored after the step were below
the limit. With the adaptive RTO the purity was below the limit only in the
four cycles of the transient, out of the 35 simulated after the move, and the
specification was recovered with a \chfour{} recovery 2.3\pp{} lower than
before the step.

After the step the plant remained at a fixed condition, so the samples
collected after it were generated at the same condition faced by the next RTO
event. In scenario 3 the disturbance continues to evolve after the
retraining.

\FloatBarrier
% -----------------------------------------------------------------------------
\subsection{Scenario 3: online learning under progressive sorbent ageing}\label{sec:caso3}
\begin{figure}[htbp]
\centering
\includegraphics[width=0.95\linewidth,height=0.86\textheight,keepaspectratio]{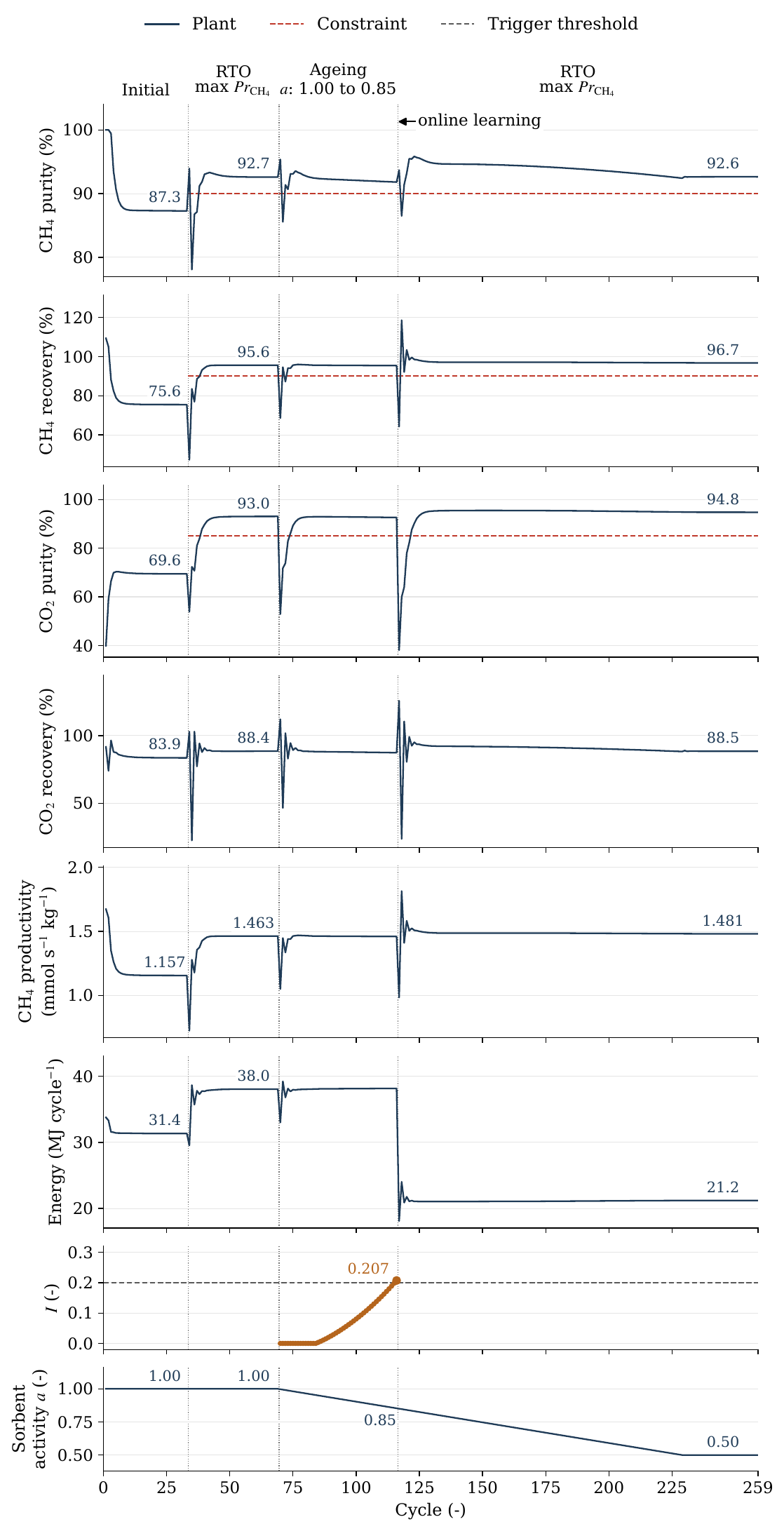}
\caption{Scenario 3: plant indicators, accumulated error $I$ and sorbent
activity $a$ in phases I to IV, with retraining at the target condition and the
objective of each RTO event above the phase. The numbers are the values at CSS
in each phase, with phase IV at the end of the decay ($a=0.50$), and the dashed
lines are the constraints and the trigger threshold. The filled circle marks
the cycle in which retraining was triggered.}
\label{fig:caso3}
\end{figure}

% -----------------------------------------------------------------------------
In scenario 3 the disturbance is a progressive loss of sorbent capacity,
which is also absent from the surrogate inputs. It also has four phases
(Fig.~\ref{fig:caso3}): (I) start-up at $\mathbf{u}_0$ with fresh sorbent; (II) one RTO event, whose CSS purity defines the reference $r$
of Eq.~\eqref{eq:integral}; (III) activity decay with the operating point of
phase II kept and the trigger monitored; and (IV) retraining, a new RTO event
and continuation of the decay to $a=0.50$, followed by 30 cycles at the final
activity. The RTO events of phases II and IV solve
\begin{equation}
\max_{\mathbf{u}}\ Pr_{\chfour}\quad\text{s.t.}\quad
P_{\chfour}\ge 90\,\%,\ R_{\chfour}\ge 90\,\%,\ P_{\cotwo}\ge 85\,\%,\
s_{\mathrm{R}}s_{\mathrm{L}}\le 0.45,
\label{eq:prob3}
\end{equation}
in the form of Problem~\eqref{eq:rto}, with the settings of scenario 1. The
trigger is that of scenario 2, with the first 15 cycles of the decay excluded
and without the persistence.

The RTO event of phase II returned a point predicted at 92.67\,\% purity, and
the plant reached 92.68\,\% after 17 cycles, which was adopted as the reference.
The activity started to decrease at the beginning of phase III. Because the
surrogate has no information on the activity, its prediction for the operating
point in force remained at 92.67\,\%, while the plant purity decreased
gradually. The accumulated error of Eq.~\eqref{eq:integral} reached the
threshold after 47 cycles of decay, at $a=0.853$, with $I=0.207$ and an
instantaneous deviation of 0.87\pp{} from the reference. The plant purity at
that moment was 91.81\,\%, above the 90\,\% limit. Had the operating point been
kept, the purity would have fallen below the limit about 65 cycles later, at
$a\approx0.65$, and would have reached 83.27\,\% at the end of the decay
(Table~\ref{tab:cond}). The loss of capacity has the same signature as the
feed step of scenario 2, at a slower rate: the bed retains less \cotwo{} per
cycle, and with the operating point kept the \cotwo{} recovery would have
decreased from 88.4 to 71.2\,\% at the end of the decay, while the \chfour{}
recovery and the productivity would have remained at 95.6\,\% and
1.463\,mmol\,s$^{-1}$\,kg$^{-1}$. The decrease in purity is not uniform along
the decay: 0.9\pp{} over the first 15\,\% of capacity lost, 2.1\pp{} over the
following 20\,\% and 6.4\pp{} over the last 15\,\%.

The threshold $I^{\star}=0.20$ corresponds, for a constant deviation, to
20 cycles at 1\,\% below the reference or 40 cycles at 0.5\,\%. In this case
it gave a lead of about 65 cycles between the trigger and the first constraint
violation that would have occurred without retraining.

Table~\ref{tab:cond} compares the outcome of the two
retraining conditions of Section~\ref{sec:cond}. In both, the local samples
were generated in two passes, 30 in a box of $\pm40\,\%$ around the operating
point of phase II and 10 in a box of $\pm10\,\%$ around the point proposed by
the first optimization, each integrated for at most 40 cycles, and the network
was fine-tuned with the settings of scenario 2. For the range condition,
$\beta$ was set to 4.27\pp, half the decrease in purity that the operating
point of phase II would undergo over the remaining decay.

\begin{table}[!htb]
\centering
\caption{Scenario 3: outcome at the end of the decay ($a=0.50$) for the
retraining conditions compared.}
\label{tab:cond}
\small
\begin{tabular}{lccccc}
\toprule
Retraining data & $t_{\mathrm{F}}$ & Predicted $P_{\chfour}$ & Plant $P_{\chfour}$ &
Minimum $P_{\chfour}$ & $Pr_{\chfour}$ \\
 & (s) & (\%) & (\%) & (\%) & (mmol\,s$^{-1}$\,kg$^{-1}$) \\
\midrule
No retraining             & 193.5 & 92.67 & 83.27 & 83.27 & 1.463 \\
Range, $a\in[0.50,0.853]$ & 162.6 & 95.35 & 88.56 & 87.86 & 1.508 \\
Target, $a=0.50$          & 109.7 & 93.28 & 92.64 & 92.42 & 1.481 \\
\bottomrule
\end{tabular}
\end{table}

With data spanning the activity range, 28 of the 30 samples of the first pass
and 38 of the 40 samples in total reached CSS. These samples come from plant
states with different capacities, and since the activity is not an input of
the network, samples with similar decision variables and different activities
correspond to the same input with different indicators, among which the
fine-tuned network can only fit a compromise. Its validation error after
fine-tuning is similar to that obtained with the target condition
(Table~\ref{tab:mae}), 0.57 against 0.51\pp{} in the \chfour{} purity, but it
is measured on samples drawn over the same range of activities and therefore
does not indicate the error at the final activity. The optimum was predicted at
95.35\,\% purity. At the plant, the purity at this point was 98.3\,\% in the
first cycles after the move, when the activity was still close to 0.85, and
decreased to 88.56\,\% at the end of the decay; the prediction lies between
these two values, and the deviation from the plant at the end of the decay was
6.79\pp{} The purity fell below the 90\,\% limit at $a\approx0.54$ and remained
below it in the last 44 cycles. The feed step of this point is shorter than
that of phase II, but it still admits 325\,mol of \cotwo{} per cycle, against
219\,mol for the point obtained with the target condition. In the last 30 cycles, at $a=0.50$, the range-trained optimum showed
cycle-to-cycle oscillations with amplitudes of 0.70\pp{} in purity and
0.045\,mmol\,s$^{-1}$\,kg$^{-1}$ in productivity, against 0.05\pp{} and
0.0002\,mmol\,s$^{-1}$\,kg$^{-1}$ for the target-trained optimum.

With data generated at the target activity, all samples represent the plant
condition at the end of the decay. In the first pass, 25 of the 30 samples
reached CSS and the network was fine-tuned for 80 epochs. After the second
pass, 35 valid samples were available and the network of phase II was
fine-tuned on them for 400 epochs; the validation errors of the purities and
recoveries decreased to at most 0.9\pp{} (Table~\ref{tab:mae}). The optimum was predicted at 93.28\,\% purity, and the
plant reached 92.64\,\% at $a=0.50$. The prediction exceeded the plant value by
0.64\pp{} During the remaining decay the
activity was higher than the value represented in the retraining data: the
plant purity rose to 95.8\,\% in the cycles after the move and then decreased
gradually to the value at the final activity. After the transient of the move, in which the purity was
below 90\,\% for one cycle, the minimum was 92.42\,\%, reached at the end of
the decay.

The feasible optimum had a productivity of 1.481\,mmol\,s$^{-1}$\,kg$^{-1}$,
about 2\,\% below the infeasible range-trained optimum and 1.2\,\% above the optimum
of phase II with fresh sorbent. The adaptation acted mainly on the feed step,
which is 43\,\% shorter (109.7 against 193.5\,s) for a loss of capacity of
50\,\%, so that the \cotwo{} admitted per cycle decreased from 387 to 219\,mol,
in line with the capacity that remains. Because the cycle is shorter in the
same proportion, the feed flow rate processed is unchanged and the
productivity is preserved, and the bed with half of its capacity is used in
shorter cycles. The energy per cycle decreased from 38.0 to 21.2\,MJ and the
specific energy from 68.6 to 66.6\,kJ\,mol$^{-1}$. Under a monotonic
disturbance with known end point, generating the retraining data at that end
point gave an operating point that remained feasible over the remaining
trajectory after the transient of the move, with a single retraining.

Table~\ref{tab:adapt} brings together the two disturbances and compares the
operation with the operating point of phase II kept, which is also the result
of an RTO without update of the model, and with the adaptive RTO.

\begin{table}[htbp]
\centering
\caption{Operation with the operating point of phase II kept and with the
adaptive RTO. Scenario 2: point kept, at the trigger; adaptive, at CSS.
Scenario 3: both at the end of the decay ($a=0.50$). The last column gives the
cycles with $P_{\chfour}<90\,\%$ after the disturbance (point kept) or after
the RTO move (adaptive), over the cycles simulated.}
\label{tab:adapt}
\small
\begin{tabular}{llcccccc}
\toprule
Scenario & Operation & $t_{\mathrm{F}}$ & $P_{\chfour}$ & $R_{\cotwo}$ &
$Pr_{\chfour}$ & $E_{\mathrm{s}}$ & Cycles below \\
 & & (s) & (\%) & (\%) & (mmol\,s$^{-1}$\,kg$^{-1}$) & (kJ\,mol$^{-1}$) & the limit \\
\midrule
2 & Point kept   & 191.5 & 76.95 & 69.4 & 1.241 & 93.4 & 20 of 21 \\
2 & Adaptive RTO & 132.1 & 90.31 & 90.0 & 1.214 & 88.2 & 4 of 35 \\
3 & Point kept   & 193.5 & 83.27 & 71.2 & 1.463 & 64.2 & 50 of 160 \\
3 & Adaptive RTO & 109.7 & 92.64 & 88.5 & 1.481 & 66.6 & 1 of 143 \\
\bottomrule
\end{tabular}
\end{table}

With the operating point kept, the light product ends 6.7\pp{} below the
specification at the end of the decay, and 50 of the 160 cycles of the decay
are below the limit, against one cycle, in the transient of the move, with the
adaptive RTO. As after the feed step, the loss is in the quality of the
product and not in its amount: the \cotwo{} that the bed no longer retains
leaves with the light product, while the \chfour{} recovery changes by less
than 0.3\pp, and the productivity at the end of the decay is the same as with
fresh sorbent, so the loss of capacity is not visible in the production rate.

In both scenarios the adaptive RTO recovered the specification with the same
type of action, a shorter feed step that reduces the \cotwo{} admitted per
cycle, by 14\,\% after the feed step and by 43\,\% under the loss of capacity,
and with a small effect on the other indicators: in scenario 3 the
productivity and the specific energy are within 3\,\% of the values with fresh
sorbent. The purity of the light product thus depends on how the \cotwo{}
admitted per cycle compares with the capacity of the bed. Feed composition and
sorbent capacity change this relation during operation, and a surrogate of the
CSS identified at nominal conditions has no information on either, so the
operating point can only be corrected if the model is updated with data from
the current condition of the plant. The progressive disturbance also shows the
role of the trigger: the loss of purity accelerates as the capacity decreases,
and the accumulated deviation started the retraining about 65 cycles before
the limit would have been reached.

\FloatBarrier
% =============================================================================
\section{Conclusions}\label{sec:conc}
% =============================================================================
An RTO framework for PSA units was developed in which a neural network
surrogate of the CSS indicators is optimized by particle swarm and updated
online by fine-tuning on samples generated at the plant. Its application to a
\chfour/\cotwo{} separation, in three scenarios with a rigorous dynamic model
as virtual plant, led to the following conclusions.

Without disturbance, five consecutive RTO events with different
objectives returned feasible optima, with deviations between surrogate and
plant of at most 1.7\pp{} The plant was on the safe
side of the prediction for all the constraints that bounded the optima.

A step in feed composition, which is not an input of the surrogate, reduced the
\chfour{} purity from 92.3 to 76.9\,\% at constant operating point, while the
surrogate prediction remained unchanged. The accumulated error crossed the
threshold 17 cycles after the step and, after remaining above it for five
cycles, triggered retraining. Fine-tuning on 43 local samples reduced the
validation error of the purity from 9.3 to 0.9\pp, and the subsequent RTO
event returned a feasible optimum with a purity of 90.3\,\%, 1.9\pp{} below
the prediction.

For a progressive loss of sorbent capacity, the accumulated relative deviation
from a purity reference triggered retraining at 85\,\% of the initial activity,
before any constraint violation and about 65 cycles before the first violation
that would have occurred without retraining. The condition at which the
retraining data were generated determined the feasibility of the result. With
data spanning the remaining activity range, the optimum reached 88.56\,\% purity
at the end of the decay, below the 90\,\% limit. With data generated at the final activity, the
optimum remained feasible over the whole decay after the transient of the
move, with a deviation of 0.64\pp{} between prediction and plant at the end.

In both disturbed scenarios the loss was in the purity of the light product,
with the \cotwo{} no longer retained by the bed leaving with it, while the
\chfour{} recovery changed by less than 0.3\pp{} The adaptive RTO restored the
specification mainly by shortening the feed step, by 31\,\% after the feed
step and by 43\,\% for a loss of capacity of 50\,\%, and in the ageing
scenario the productivity and the specific energy remained within 3\,\% of
the values with fresh sorbent.

Future work will address measurement noise, the estimation of the end-point
condition when it is not known in advance, and a transition strategy that
limits the transient violations after each RTO move.

% =============================================================================
\section*{CRediT authorship contribution statement}
\textbf{Carine de Menezes Rebello:} Conceptualization, Methodology, Software,
Validation, Formal analysis, Investigation, Data curation, Visualization,
Writing - original draft.
\textbf{Marcio Fontana:} Writing - review \& editing.
\textbf{Leizer Schnitman:} Writing - review \& editing.
\textbf{Idelfonso B. R. Nogueira:} Conceptualization, Methodology, Resources,
Supervision, Project administration, Writing - review \& editing.

\section*{Declaration of competing interest}
The authors declare that they have no known competing financial interests or
personal relationships that could have appeared to influence the work reported
in this paper.

\section*{Data availability}
Data will be made available on request.

\section*{Acknowledgements}
The authors acknowledge the support of Puffin HUB.

% =============================================================================
\appendix
\section{Rigorous model of the PSA column}\label{sec:model}
% =============================================================================
The virtual plant is a one-dimensional, non-isothermal model of a packed
column. It adopts the assumptions of the PSA model of our previous work
\cite{rebello2024}: ideal gas behaviour; no radial gradients of concentration,
pressure or temperature; axially dispersed plug flow; adsorbent properties and
bed porosity constant along the column; thermal equilibrium between the gas
and the adsorbent; mass transfer described by a linear driving force (LDF)
rate \cite{glueckauf1955}; and pressure drop given by the Ergun equation.
Differently from that model, the equations are written here in dimensional
form and an energy balance for the column wall is included.

The mass balance of component $i$ in the gas phase is
\begin{equation}
\varepsilon\,\frac{\partial C_i}{\partial t}
=-\frac{\partial (u\,C_i)}{\partial z}
+\varepsilon\,D_{\mathrm{ax}}\frac{\partial^{2} C_i}{\partial z^{2}}
-(1-\varepsilon)\,\rho_{\mathrm{p}}\frac{\partial q_i}{\partial t},
\label{eq:massbal}
\end{equation}
where $C_i$ is the gas concentration, $q_i$ the adsorbed amount, $u$ the
superficial velocity, $\varepsilon$ the bed porosity, $\rho_{\mathrm{p}}$ the
particle density and $D_{\mathrm{ax}}$ the axial dispersion coefficient. The
adsorption rate is given by the LDF model,
\begin{equation}
\frac{\partial q_i}{\partial t}=k_i\left(q_i^{*}-q_i\right),
\qquad
k_i=\frac{15\,D_{\mathrm{p},i}}{R_{\mathrm{p}}^{2}}\,\frac{P_{\mathrm{H}}}{P},
\label{eq:ldf}
\end{equation}
where $D_{\mathrm{p},i}$ is the diffusivity in the particle, $R_{\mathrm{p}}$
the particle radius, $P$ the local pressure and $P_{\mathrm{H}}$ the high
pressure of the cycle. The adsorbed amount at equilibrium, $q_i^{*}$, follows
a Langmuir isotherm with temperature-dependent affinity,
\begin{equation}
q_i^{*}=\frac{a\,q_{\mathrm{sat},i}\,b_i\,p_i}{1+\sum_j b_j\,p_j},
\qquad
b_i=b_{0,i}\exp\!\left(\frac{-\Delta H_i}{R\,T}\right),
\label{eq:iso}
\end{equation}
where $p_i=C_iRT$ is the partial pressure and $a\in(0,1]$ is the sorbent
activity introduced in Section~\ref{sec:dist}, equal to 1 for fresh sorbent.
The pressure drop along the bed is given by the Ergun equation,
\begin{equation}
-\frac{\partial P}{\partial z}
=\frac{150\,\mu\,(1-\varepsilon)^{2}}{\varepsilon^{3}d_{\mathrm{p}}^{2}}\,u
+\frac{1.75\,(1-\varepsilon)}{\varepsilon^{3}d_{\mathrm{p}}}\,\rho_{\mathrm{g}}\,u\,|u|,
\label{eq:ergun}
\end{equation}
where $\mu$ is the gas viscosity, $d_{\mathrm{p}}$ the particle diameter and
$\rho_{\mathrm{g}}$ the gas density.

With $C=\sum_i C_i$ and the gas and the adsorbent at the same temperature $T$,
the energy balance of the bed is
\begin{equation}
\begin{aligned}
(\rho c)_{\mathrm{b}}\,\frac{\partial T}{\partial t}
={}&-\frac{\partial (u\,C\,C_{p,\mathrm{g}}\,T)}{\partial z}
+\lambda\,\frac{\partial^{2} T}{\partial z^{2}}
+(1-\varepsilon)\,\rho_{\mathrm{p}}\sum_i(-\Delta H_i)\frac{\partial q_i}{\partial t}\\
&+\left[\varepsilon+(1-\varepsilon)\,\varepsilon_{\mathrm{p}}\right]R\,T\,\frac{\partial C}{\partial t}
-\frac{4\,h_{\mathrm{w}}}{D}\left(T-T_{\mathrm{w}}\right),
\end{aligned}
\label{eq:energy}
\end{equation}
in which the heat capacity of the bed per unit volume is
$(\rho c)_{\mathrm{b}}=\varepsilon\,C\,C_{v,\mathrm{g}}+(1-\varepsilon)
\big[\varepsilon_{\mathrm{p}}\sum_i C_i\,C_{v,i}+\rho_{\mathrm{p}}\sum_i
q_i\,C_{v,i}+\rho_{\mathrm{p}}\,C_{p,\mathrm{s}}\big]$. The terms on the right
are convection, axial conduction, the heat of adsorption, the effect of the
change in gas concentration and the exchange with the wall; $C_{p,\mathrm{g}}$
and $C_{v,\mathrm{g}}$ are the heat capacities of the gas mixture, $C_{v,i}$
that of component $i$, $C_{p,\mathrm{s}}$ that of the adsorbent,
$\varepsilon_{\mathrm{p}}$ the particle porosity, $\lambda$ the axial thermal
conductivity, $h_{\mathrm{w}}$ the heat transfer coefficient between the gas
and the wall and $D$ the column diameter. The wall temperature
$T_{\mathrm{w}}$ follows
\begin{equation}
\rho_{\mathrm{w}}\,C_{p,\mathrm{w}}\,\frac{\partial T_{\mathrm{w}}}{\partial t}
=\alpha_{\mathrm{w}}\,h_{\mathrm{w}}\left(T-T_{\mathrm{w}}\right)
-\alpha_{\mathrm{wl}}\,U\left(T_{\mathrm{w}}-T_{\infty}\right),
\label{eq:wall}
\end{equation}
where $\rho_{\mathrm{w}}$ and $C_{p,\mathrm{w}}$ are the density and heat
capacity of the wall, $\alpha_{\mathrm{w}}$ and $\alpha_{\mathrm{wl}}$ the
specific areas of the wall for the exchange with the gas and with the
surroundings, $U$ the overall heat transfer coefficient to the surroundings
and $T_{\infty}$ the ambient temperature, taken equal to the feed temperature.

The velocity and pressure profiles depend on the step. In the steps with flow
through the column at constant outlet pressure (feed, rinse and purge), the
molar flux at the inlet is specified, the flux along the bed follows from the
overall mass balance and the pressure profile from Eq.~\eqref{eq:ergun},
integrated from the outlet. In the blowdown the inlet is closed and the outlet
pressure decays exponentially from the high to the low pressure. In the
pressurization the outlet is closed, the pressure follows from the ideal gas
law and the velocity from Eq.~\eqref{eq:ergun}, and the step ends when the
high pressure is reached. At an inlet, the composition and the temperature of
the entering stream are imposed; at an outlet, the dispersive and conductive
fluxes are zero. The main parameters are listed in Table~\ref{tab:param}.

The equations are discretized in space by finite volumes with 50 cells and
first-order upwind convection, and the resulting system of ordinary
differential equations is integrated with the KenCarp4 method of
DifferentialEquations.jl \cite{rackauckas2017,kennedy2003} with relative and
absolute tolerances of $10^{-6}$ and $10^{-8}$. The cycle is simulated for a
single column, with the streams exchanged between steps stored and reused: the
rinse step receives the heavy product of the previous cycle, while purge and
pressurization receive the light product of the current one. The column state
at the end of each cycle is the initial state of the next, and when the
operating point changes the simulation continues from the current state.

\begin{table}[htbp]
\centering
\caption{Column, sorbent and isotherm parameters of the virtual plant.}
\label{tab:param}
\small
\begin{tabular}{lll}
\toprule
Parameter & Value & Unit \\
\midrule
Column length; diameter & 1.5; 0.9 & m \\
Sorbent mass & 490 & kg \\
Bed porosity; particle porosity & 0.38; 0.41 & - \\
Particle density; particle radius & 828; 1.5 & kg\,m$^{-3}$; mm \\
$q_{\mathrm{sat}}$ (\cotwo; \chfour) & 6.377; 7.381 & mol\,kg$^{-1}$ \\
$b_0$ (\cotwo; \chfour) & $1.55\times10^{-5}$; $8.07\times10^{-5}$ & bar$^{-1}$ \\
$-\Delta H$ (\cotwo; \chfour) & 26.4; 18.3 & kJ\,mol$^{-1}$ \\
$D_{\mathrm{p}}$ (\cotwo; \chfour) & $6.4\times10^{-7}$; $8.93\times10^{-7}$ & m$^{2}$\,s$^{-1}$ \\
Axial dispersion coefficient & $2.6\times10^{-6}$ & m$^{2}$\,s$^{-1}$ \\
$h_{\mathrm{w}}$; $U$ & 60; 40 & W\,m$^{-2}$\,K$^{-1}$ \\
$\alpha_{\mathrm{w}}$; $\alpha_{\mathrm{wl}}$ & 1.69; 4.4 & m$^{-1}$ \\
High pressure; low pressure & 5.0; 0.1 & bar \\
Feed temperature & 303.15 & K \\
\bottomrule
\end{tabular}
\end{table}

% =============================================================================
\bibliographystyle{model1-num-names}
\bibliography{refs}

\end{document}